# Coherent Optical Coupling to Surface Acoustic Wave Devices

Arjun Iyer[1], Yadav P. Kandel[2], Wendao Xu[1], John M. Nichol[2] and William H. Renninger[1,2]

[1] Institute of Optics, University of Rochester, Rochester, NY 14627, USA.
[2] Departament of Physics and Astronomy, University of Rochester, Rochester, NY 14627, USA.
*email: aiyer2@ur.rochester.edu

## Abstract

Surface acoustic waves (SAW) and associated SAW devices are ideal for sensing, metrology, and connecting and controlling hybrid quantum devices. While the advances demonstrated to date are largely based on electromechanical coupling, a robust and customizable coherent optical coupling would unlock mature and powerful cavity optomechanical control techniques and an efficient optical pathway for long-distance quantum links. Here we demonstrate direct and robust coherent optical coupling to surface acoustic wave cavities through a Brillouin-like optomechanical interaction. In high-frequency SAW cavities designed with curved metallic acoustic reflectors deposited on crystalline substrates, SAW modes are efficiently optically accessed in piezo-active directions that can be accessed through traditionally electromechanical techniques as well as non-piezo-active directions that cannot. The non-contact nature of the optical technique enables controlled analysis of dissipation mechanisms and access to pristine mechanical resonators with record-level quality factors (>$10^5$ measured here). The exceptional control of the optical probe beams also enables detailed transverse spatial mode spectroscopy, for the first time. These advantages combined with simple fabrication, small size, large power handling, and strong coupling to quantum systems make SAW optomechanical platforms particularly attractive for sensing, material science, and hybrid quantum systems.

## Main

Surface acoustic wave devices based on electrically driven piezoelectric materials are essential to modern technologies, including for communications[1–3] and chemical and biological sensors[4,5]. SAWs have more recently emerged as an exciting resource for quantum systems [6–9] because of their low loss, tight surface confinement, and strong coupling to a variety of quantum systems. As "universal quantum transducers,"[6,7] SAWs and associated manipulation and probing techniques have been demonstrated in color centers[10], superconducting qubits[11–14], semiconductor quantum dots[15–17], 2D materials[18–20], and superfluids[21,22]. While electrical control of SAWs has matured over the past several decades[1], robust and customizable coherent optical coupling has not yet been demonstrated. Coherent optical coupling would enable powerful techniques established in cavity optomechanical systems, such as quantum transduction[23,24], quantum-limited force and displacement sensing[25–29], generation of non-classical states of optical and acoustic fields[30–32] and ground state cooling of mechanical resonators[33–36]. In addition, robust optomechanical coupling to SAW devices would enable an ideal optical pathway for long-distance quantum links, a longstanding goal of experimental quantum information science[37–39].

While optical surface Brillouin scattering has been successful for probing incoherent thermal surface phonons for the study of thin films[40,41], many optomechanical applications, including for quantum science, require coherent and stimulated interactions. Brillouin processes have recently enabled coherent coupling to bulk acoustic modes with record lifetimes in shaped crystals, but with the low optomechanical coupling associated with larger bulk mode volumes[42,43]. In nanoscale systems designed for cavity optomechanics, large coupling strengths are available, but often with complex designs, including sub-wavelength and suspended structures which can be challenging to integrate into larger hybrid quantum devices. In addition nanoscale confinement also leads to undesirable heating effects[9,43,44], which limit photon numbers and acoustic quality factors, and can require complex optical pumping schemes [45,46]. SAW devices combine intrinsically tight confinement on the surface of bulk substrates with the potential for high power handling and simple fabrication. In recent SAW-based microwave-to-optical transduction schemes[9,23,47,48], electrically generated SAWs are coupled to acoustic modes of a distinct resonator, such as a nanomechanical phononic crystal cavity. While demonstrating exceptional optomechanical coupling strengths, these devices achieve low net efficiencies, primarily limited by the phonon injection efficiency of the electrically generated SAWs to the acoustic cavity modes[7,49,50]. In an alternative approach, Okada et.al[51] examined cavity optomechanical systems mediated directly with SAWs. However, without optomechanical phase matching, efficient SAW confinement, and modal size matching between optical and acoustic fields, the system is limited to lower phonon frequencies, mechanical quality factors, and coupling strengths. Optomechanical coupling to SAW whispering gallery modes in microresonators[52,53] and SAWs driven by optical absorption and thermal relaxation of metallic electrodes[54,55] also enable several new possibilities. However, these devices will be challenging to integrate with a range of qubit systems, which require specific geometries and minimal heating. Direct, efficient, coherent optical access to simple high-quality and high-power handling SAW devices will be an important step toward realizing the full promise of classical and quantum SAW-based technologies.

Here we establish a frequency tunable Brillouin-like optomechanical coupling with integrable surface acoustic wave devices that is direct, coherent, power-tolerant, and efficient. The technique is demonstrated with simple single-crystalline substrates supporting long-lived Gaussian SAW cavity modes confined with deposited curved metallic grating mirrors. Strong optomechanical coupling is demonstrated by engineering phase-matched Brillouin-like interactions between the trapped acoustic modes and incident out-of-plane non-collinear optical fields. In contrast to previous studies on SAW resonators, the demonstrated technique does not require piezoelectricity and can be applied to practically any crystalline media, which we demonstrate by optically driving piezo-inactive SAW devices. This approach, therefore, enables access to high-Q surface acoustic modes on quantum-critical materials which are not piezo-active, such as diamond and silicon. In addition, the absence of interdigital transducers or any other acousto-activating device enables cavity-limited quality-factors, including the record ~120,000 quality-factors demonstrated in GaAs in this report. The presented cavities operating at 500 MHz can be tuned through several GHz by varying the incident angle of the optical beams through simple phase matching (momentum conservation) considerations. The frequency and material versatility of this coupling technique is well suited for materials spectroscopy, as illustrated here through measuring and characterizing phonon dissipation mechanisms in GaAs cavities with metallic mirrors. The SAW optomechanical platform presented combines the simplicity and power-handling advantages of bulk optomechanical systems with the small acoustic mode volumes of nanoscale systems for enhanced interaction strengths enabling a multi-functional integrated platform for sensing, quantum processing, and condensed matter physics.

# Non-Collinear Brillouin-like Optical Coupling to Surface Acoustic Waves

The SAW device consists of a Fabry-Perot Gaussian surface acoustic wave cavity on a single-crystalline substrate formed by two acoustic mirrors composed of regularly spaced curved metallic reflectors. Two non-collinear optical beams, a pump field, and a Stokes field, are incident in the region enclosed by the acoustic mirrors (Fig. 1a). The confined Gaussian surface acoustic mode can mediate energy transfer between the two optical fields provided phase-matching (momentum conservation) and energy conservation relations are satisfied, as is the case with Brillouin scattering from bulk acoustic waves[56]. For pump and Stokes fields with wavevector (frequency) $\vec{k_p}(\omega_p)$ and $\vec{k_s}(\omega_s)$, respectively, that subtend equal but opposite angles, $\theta$, with respect to the surface normal (z-axis), the optical wavevector difference can be approximated as $\overrightarrow{\Delta k} \approx 2k_0 \sin\theta \ \hat{x}$, assuming $k_p \approx k_s = k_0$ and for $\hat{x}$ a unit vector parallel to the surface; the corresponding optical frequency difference is $\Delta\omega = \omega_p - \omega_s$ (Fig. 1b). Note that the magnitude of the optical wavevector difference is tunable by the optical angle of incidence. For the case of freely propagating surface acoustic waves, the acoustic dispersion relation is linear and can be expressed as $\Omega = qv_R$, where $\Omega$, $q$, and $v_R$ are the phonon frequency, phonon wavevector, and Rayleigh SAW velocity, respectively. The phase-matched phonon wavevector ($q_0$) and frequency ($\Omega_0$) are then given by the relations: $q_0 = \Delta k = 2k_0 \sin\theta$ and $\Omega_0 = q_0 v_R$. A propagating SAW, therefore, yields a single-frequency optomechanical response, similar to the standard Brillouin response in bulk materials from propagating longitudinal waves. However, the accessible phonon spectrum is significantly modified in the presence of a surface acoustic cavity and optical beams with finite beam sizes, as illustrated by the modified acoustic dispersion plot in Fig. 1c. First, because standing SAW cavity modes are formed, the phonon wavevectors and frequencies become discretized to specific values $q_m = \frac{m\pi}{L_{\text{eff}}}$ and $\Omega_m = q_m v_R$, respectively, characterized by mode number $m$, where the free spectral range of the cavity is $\Delta\Omega = \frac{\pi v_R}{L_{\text{eff}}}$, and $L_{\text{eff}}$ is the effective cavity length. Second, unlike ideal mirrors, acoustic Bragg mirrors only efficiently confine a finite number of longitudinal modes (blue circles in Fig. 1c) determined by the reflectance and periodicity of the metallic reflectors[1]. Finally, because the optical fields are Gaussian beams with finite spatial extents, appreciable optomechanical coupling exists over a range of optical wavevectors values centered around the phase-matched configuration, $\Delta k = q_m$. The effective optomechanical coupling rate to the cavity mode $m$, $g_0$, varies as a function of optical wavevector mismatch as $g_0(\Delta k) \propto \exp(-(\Delta k - q_m)^2/\delta k^2)$, where $\delta k = 2\sqrt{2}/r_0$ and $r_0$ is the radius of incident optical fields. Equivalently, the coupling rate can be expressed as a function of the angle of incidence as $g_0(\theta) \propto \exp(-(\theta - \theta_m)^2/\delta\theta^2)$, for small angles such that $\sin\theta \approx \theta$ and where $\theta_m$ is the phase-matching angle of the acoustic cavity mode given by $\theta_m = \frac{q_m}{2k_0}$. The corresponding angular bandwidth is given as $\delta\theta = \frac{\delta k}{k_0} = \frac{\sqrt{2}}{r_0 k_0}$ (see Section S2 of supplementary information). The resultant optomechanical spectrum consists of several discrete resonances from SAW cavity modes which lie both within the acoustic mirror and optical phase-matching bandwidths (unconfined, radiative, longitudinal modes are indicated by grey circles in Fig. 1c).

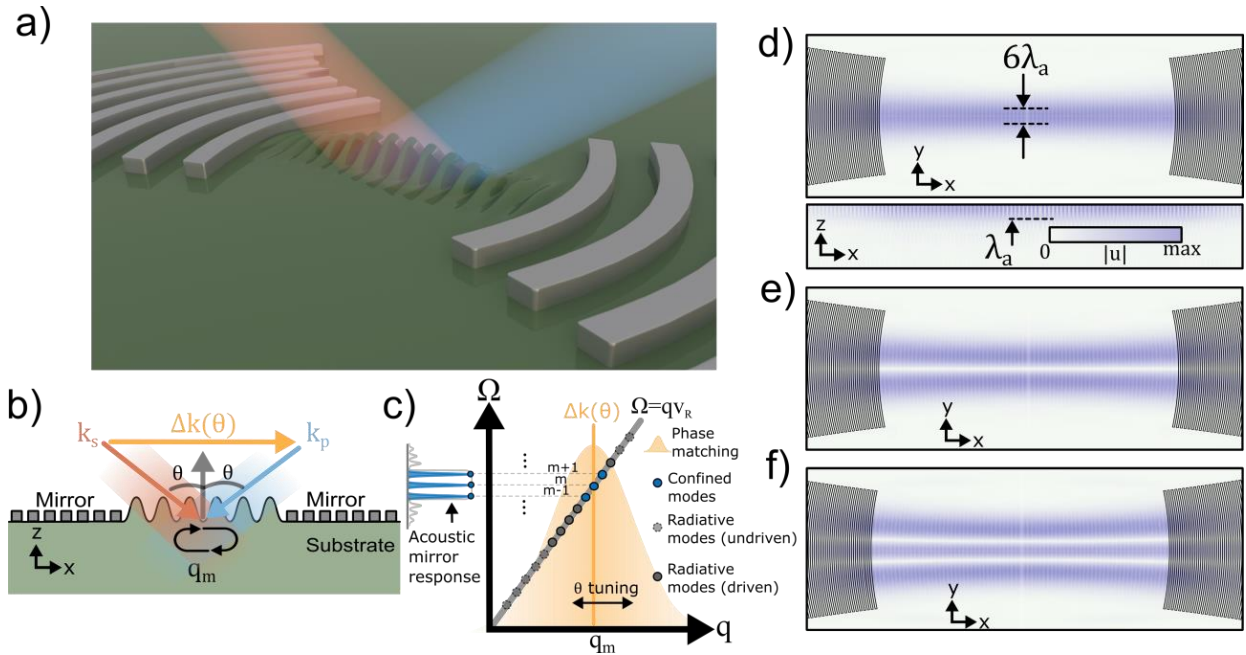

Figure 1. Parametric optomechanical interactions mediated by Gaussian SAW resonators. a) Two non-collinear traveling optical fields are incident on a Fabry-Perot type Gaussian SAW resonator; interaction between the two optical fields is mediated by a Gaussian SAW cavity mode confined to the surface of the substrate. b) Phase-matching diagram of the parametric process. The vectorial optical wavevector difference, $\Delta\vec{k} = \vec{k_p} - \vec{k_s} = 2k_0 \sin\theta \ \hat{x}$, is angle-dependent and points along the direction of SAW cavity axis. c) The acoustic dispersion relation $\Omega(q)$ is discretized in the presence of a SAW cavity. The final optomechanical response is determined by the modes, which are both within the phase-matching and the acoustic mirror bandwidth (blue dots), while radiating longitudinal modes excluded by the acoustic mirror and the optical phase matching (grey dots) do not yield an optomechanical response. d) Finite element calculation of the acoustic displacement magnitude, $|u|$, in a SAW cavity along [100] direction on [100] GaAs illustrating the Gaussian mode (upper panel) with the designed acoustic waist of $w_a = 3\lambda_a$ and an approximate penetration depth of $\sim \lambda_a$ (lower panel). Panels e) and f) display YX cross-sections of acoustic displacement for e) anti-symmetric and f) symmetric higher-order transverse modes of the SAW cavity.

Gaussian SAW cavities are designed to achieve small acoustic mode volume and appreciable coupling strengths (see Methods and section S3 of supplementary information). Diffraction losses are mitigated by accounting for the anisotropy of the acoustic group velocity on the underlying crystalline substrate[57]. GaAs is chosen because of its large photoelastic response, ease of fabrication, and integration with other quantum systems such as qubits. 3-dimensional numerical finite element simulations are performed of a SAW cavity on [100]-cut GaAs oriented along [100]-direction. The acoustic wavelength of $\lambda_a = 5.7\ \mu m$ and Gaussian waist, $w_a = 3\lambda_a$, are near-identical to the experimental devices described below, while the number of reflectors and the mirror spacing are reduced to maintain computational feasibility (see section S3 of supplementary information). The simulated cavities display a series of stable SAW cavity modes with Hermite-Gaussian-like transverse profiles (Fig. 1d-f) separated by the free spectral range of the cavity. As expected, the modes are confined to the surface and steeply decay into the bulk of the substrate (e.g. lower panel Fig. 1d). The observed beam waist of the fundamental Gaussian mode agrees well with the designed full-waist ($2w_a$) of $6\lambda_a$. Higher-order anti-symmetric (Fig. 1e) and symmetric (Fig. 1f) mode solutions are also observed.

## Optomechanical spectroscopy of SAW cavities

To demonstrate coherent optical coupling to SAW devices, Gaussian SAW cavities are fabricated (see Methods and section S4 of supplementary information) on a single crystal GaAs substrate (inset Fig. 2a).

Optomechanical measurements are made for two sets of cavities, one oriented along the crystalline [110] direction, which is piezo-active, and one along the [100] direction, which is piezo-inactive. The cavities are designed for an acoustic wavelength of $\lambda_a \approx 5.7\ \mu m$ ($\theta \approx 7.8°$), acoustic waist of $w_a \approx 4\lambda_a$ and mirror spacing of $L \sim 500\ \mu m$. The cavity parameters are chosen to optimize for practical constraints including finite optical apertures, electronics bandwidths, and the optical beam sizes. The effective cavity length ($L_{\text{eff}}$) is calculated to be ~ $610\ \mu m$ by accounting for the penetration depth into the mirrors[1,58]. The large mirror separation relative to the optical beam size minimizes absorptive effects arising from spatial overlap of optical fields with acoustic metallic reflectors (See Section S9 of supplementary information).

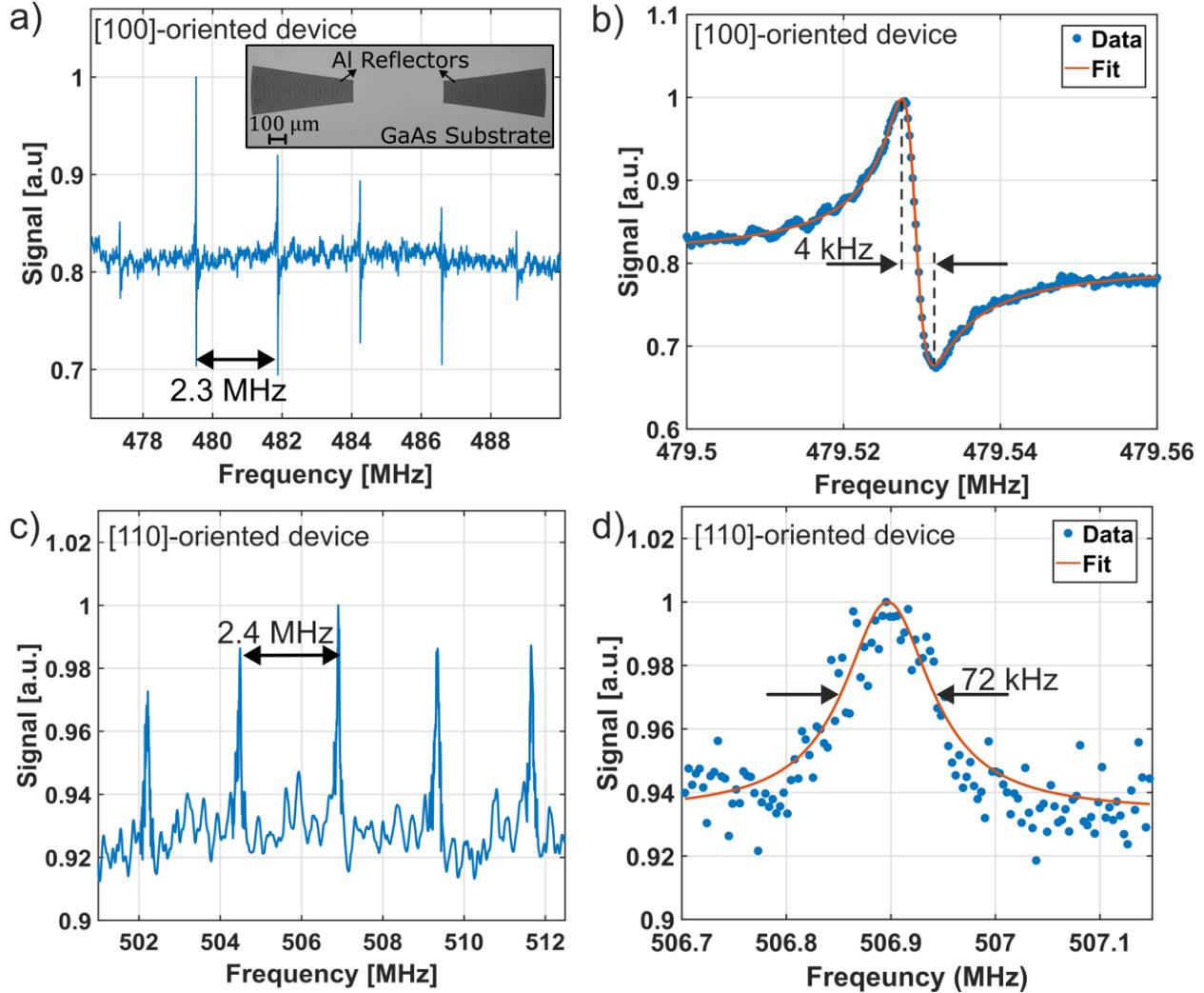

Figure 2. Optically measured SAW devices. Optomechanical response of 470 MHz SAW cavities oriented along the [100]-direction on [100]-cut GaAs at 4K with a) a wide frequency sweep revealing several discrete SAW cavity resonances separated by the 2.3 MHz cavity free spectral range (microscope image of the SAW device is inset), and b) a high-resolution frequency sweep revealing an acoustic Q-factor of 120,000 or a spectral linewidth of ~4 $kHz$. c) and d) show similar measurements for the SAW cavity oriented along [110]-direction on [100]-cut GaAs. c) A wide scan reveals SAW cavity modes separated by the free spectral range of 2.4 MHz and d) the SAW modes on the [110]-oriented devices exhibit a maximum quality factor of 7000 with a corresponding linewidth of 72 kHz.

The stimulated optomechanical response is measured with the sensitive phonon-mediated four-wave mixing measurement technique described in Methods and section S5 of supplementary information. The spectral response of the piezo-inactive (active) cavity along the [100]([110])-direction in Fig. 2a-b (c-d) reveals several equally spaced resonances over a wide spectral range centered at 480 MHz (506 MHz) separated by 2.3 MHz (2.4 MHz), which corresponds to the free-spectral range ($v_R/2L_{\text{eff}}$) of the SAW cavity. The observed resonances span $\sim 10\ MHz$, which is consistent with the designed acoustic mirror bandwidth. High-resolution spectral analysis of one of the observed SAW cavity resonances of the piezo-inactive (active) cavity (Fig. 2b(d)) reveals a spectral width, $\Gamma/2\pi$, of 4 kHz (72 kHz), corresponding to an acoustic quality factor of 120,000 (7000). The measured traveling-wave zero-point coupling rate[59,60], $g_0$, for the piezo-inactive(active) cavity of $2\pi \times 1.4\ kHz$ ($2\pi \times 1.7\ kHz$) is consistent with predicted values of $2\pi \times 1.9\ kHz$ ($2\pi \times 1.8\ kHz$) obtained using known material parameters in conjunction with the device geometry (see section S1 and section S7 of supplementary information). As expected, no measurable acoustic response is observed when either of the optical drive tones is turned off. Additionally, as predicted by theoretical coupling calculations (see section S1 of supplementary information), no resonance is observed when the acoustic drives are orthogonally polarized to each other, or when the LO is orthogonally polarized to the incident probe (see section S11 of supplementary information). The demonstrated acoustic quality factors of the SAW devices are among the highest measured for focused SAW cavities on any substrate, corresponding to an $fQ$ product of $6 \times 10^{13}$ $Hz$, which is also comparable to that of the best electromechanical SAW devices [15,58,61]. Moreover, the accessed SAW cavity modes are along electromechanically inaccessible directions, demonstrating a key merit of the coherent optical coupling in enabling access to long-lived SAW modes regardless of their piezoelectric properties. The larger relative loss in the [110]-oriented cavities is consistent with excess ohmic loss from the metallic reflectors owing to non-uniform strain from the Gaussian modes and the resulting piezoelectric potential on the reflectors[62–64].

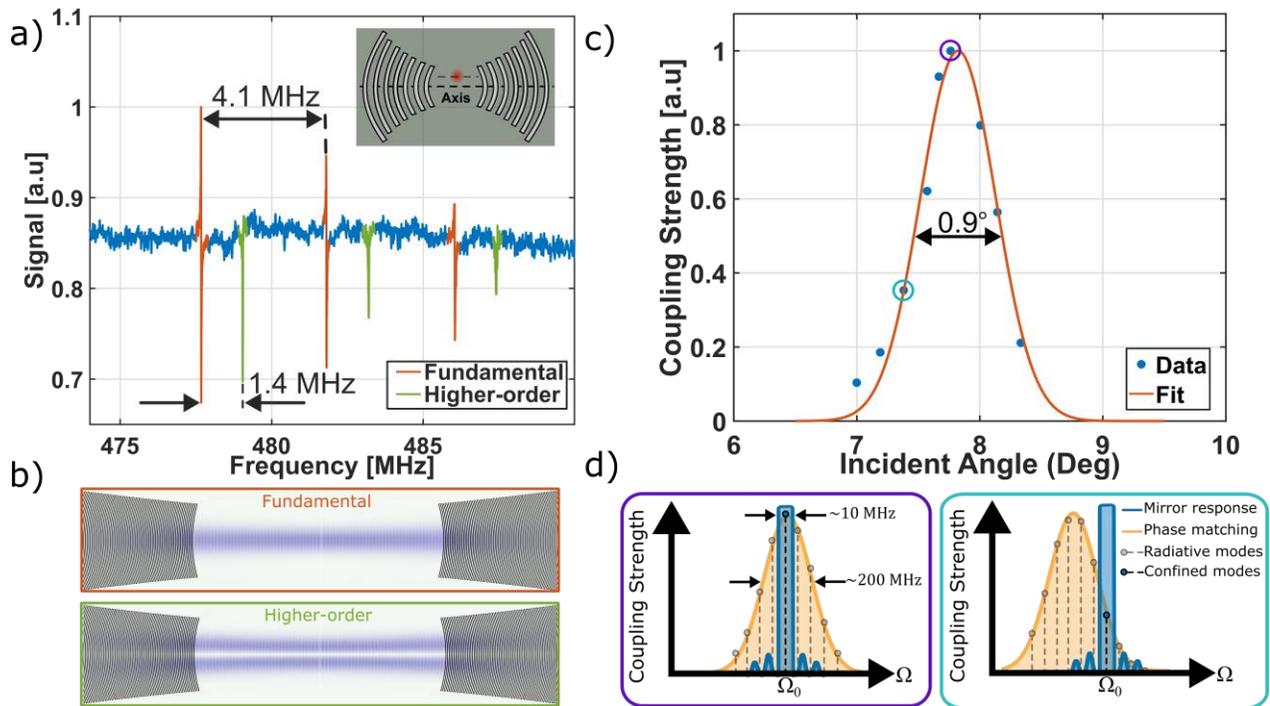

Figure 3. Measured spatial mode spectrum and angular dependence. a) The optomechanical response from higher-order acoustic modes is observed in a [100]-oriented cavity with a mirror spacing of $L \approx 350\ \mu m$ by laterally displacing the optical fields as illustrated in the inset figure. The higher-order frequency spacing of 1.4 MHz, is consistent the theoretical estimate. b) Finite element simulations of the corresponding acoustic mode profiles. c) Optomechanical coupling strength as a function of angle of incidence with a Gaussian fit overlayed. d) A graphical representation of the positions of the phase-matching envelope relative to the acoustic mirror response (not to scale) for illustrative angles of incidence indicated with the same color outline as for the respective points in c). Left: The phase-matching envelope coincides with the mirror response for maximal optomechanical coupling strength. Right: The phase matching envelope is detuned from the mirror-defined SAW mode resulting in a weaker optomechanical response.

While the high-order spatial modes of a Gaussian SAW resonator are challenging to probe electromechanically, the coherent optomechanical technique allows for precise and direct excitation of spatial modes through fine control of the optical spatial overlap with specific acoustic mode profiles. By laterally displacing the optical beams away from the cavity axis, optomechanical coupling to a specific higher-order SAW cavity mode is observed (green in Fig. 3a-b) in addition to the response from the fundamental mode (red in Fig. 3a-b). The frequency separation between the fundamental and corresponding higher-order mode of $1.4$ $MHz$ is consistent with the predicted difference of 1.4 MHz. The exquisite spatial control available through optical techniques could form the basis of novel SAW-based spatially resolved sensing and metrology.

Finally, the accessible phonon-mode bandwidth determined by phase matching is characterized through measurements of the Brillouin coupling coefficient ($G_B \propto |g_0|^2$) as a function of the angle of incidence of the optical fields (see section S2 of supplementary information). The coupling strength exhibits a Gaussian dependence on the angle (Fig. 3c) for peak coupling centered at $\theta_0 = 7.8°$, with an angular bandwidth of $0.9°$, which agrees with the predicted bandwidth of $0.96°$. The peak coupling at $\theta_0 = 7.8°$ is determined by the angle of incidence of the optical fields and the resultant wavevector difference, but the acoustic mode frequencies are independently fixed by the cavity geometry and the acoustic mirror response. The effective optomechanical coupling rate is maximized when the center of the optical phase matching envelope coincides with the peak reflection frequency of the acoustic mirrors (point outlined with purple circle in Fig. 3c and illustrated in the left panel of Fig. 3d) and decreases as they are mismatched (cyan circle in Fig. 3c and illustrated in right panel of Fig. 3d). Because the optomechanical gain bandwidth and associated driven acoustic modes result from the optical spatial profiles, this technique presents the unique capability of tailoring the optomechanical gain profile for specific applications, from multi-mode optomechanics to tunable single frequency applications.

# Non-contact Probing of SAW Cavity Dissipation Mechanisms

Acoustic dissipation is typically measured using electromechanical techniques, which also include the dissipation from external device structures such as the electrodes, electrical ports, and impendence matching circuits, limiting insights into material and structural dissipation mechanisms. In contrast, the coherent optical interaction is contact-free and not limited by these extrinsic effects. A direct probe into phonon loss mechanisms will be valuable for basic material science as well as for optimizing novel SAW device technologies. Here the coherent optical technique is used to determine the dominant loss mechanisms between SAW propagation and mirror losses for Gaussian resonators and to extract the temperature dependence of the dissipation. The acoustic quality factor is measured as a function of mirror separation (i.e. cavity length) and temperature for cavities in both the [100]-oriented (piezo-inactive) and [110]-oriented (piezo-active) cavities. The measured cavities are all designed to have identical parameters except for the mirror separation, which varies from $150 \mu m$ to $500\ \mu m$. The cavity lengths are chosen to minimize effects resulting from optical absorption in the metallic reflectors (See Section S9 of supplementary information). For both cavity orientations, the acoustic quality factor displays a linear dependence on cavity length (Fig. 4a, Fig. 4b), with the quality factor increasing for larger cavity lengths. A linear dependence on cavity length suggests that the losses in SAW cavities primarily occur within the acoustic mirrors through mechanisms such as scattering into the bulk, ohmic losses, and acoustic losses within the reflectors (see section S8 of supplementary information).

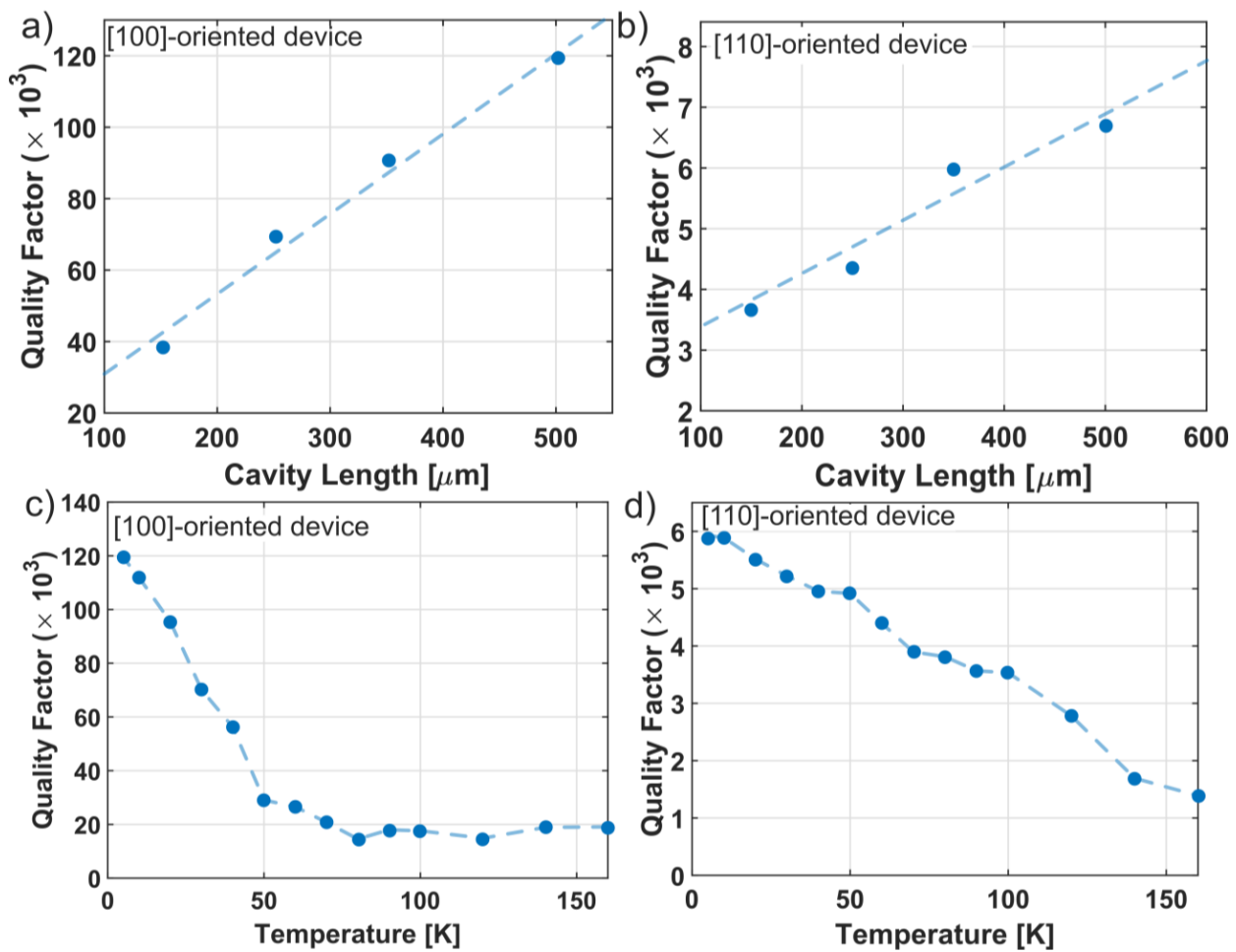

Figure 4. Characterizing loss in SAW devices. Quality factor as a function of cavity length for cavities oriented along the a) [100] and b) [110] directions. Both sets of cavities display a linear dependence of quality factor on cavity length indicating that cavity loss is dominated by the acoustic mirrors. Quality factor as a function of temperature for cavities oriented along the c) [100] and d) [110] direction. The two orientations display qualitatively distinct dependencies, suggesting differences in acoustic loss mechanisms.

The dependence of quality factor on temperature is also investigated from $T = 4$ to 160 K (Fig.4c-d) for a fixed mirror separations ($L = 500\ \mu m$ for the [100]-oriented cavities and $L = 350\ \mu m$ for the [110]-oriented cavities). The [100]-oriented cavities exhibit a sharp fall and a subsequent plateau at $Q \sim 20{,}000$ within the measured temperature range. In contrast, the [110] cavities exhibit a linear decrease of $Q$ with temperature. These measurements suggest that while both cavity types are limited by losses occurring within the acoustic mirrors, the specific mirror loss mechanisms likely differ. Previous measurements of Gaussian SAW cavities on GaAs without the potential for ohmic losses in the mirrors (through superconducting reflectors[65] as well as non-metallic reflectors[61]) demonstrating large acoustic quality factors ($2 \times 10^4$), suggest that the losses observed in the [110]-oriented, piezo-active, cavities primarily result from ohmic losses within the metallic reflectors. This is also consistent with the observations that the [100]-oriented cavities without piezoelectricity support much higher quality factors and have a distinct temperature dependent behavior. Additional insights could be derived from temperature dependent quality factor measurements at additional cavity lengths including longer lengths where the effects of mirror loss are reduced, from cavities where ohmic losses are reduced such as superconducting-mirror cavities, as well as from alternative cuts and material types. Importantly, because of the non-contact nature of coherent optical coupling, these measurements directly reflect intrinsic device properties, as opposed to details of the probe, providing a rich source of information across a wide range of relevant SAW device parameters. This is illustrated in Section S9 of supplementary information as well where the effects of optical absorption are clearly delineated from those of electrostriction through controlled measurements.

## Discussion and Conclusion

This report introduces a powerful new coherent optomechanical platform in which two non-collinear optical fields parametrically couple through surface acoustic modes of Gaussian SAW cavities. The platform offers high power handling capabilities, requires minimal fabrication, and enables a contact-free piezo-electricity independent coupling to SAW devices enabling record-high quality factor devices in GaAs crystalline substrates. From the results presented here there are several directions in which specific metrics of interest for applications can be improved. For example, the principles outlined here can be readily applied for the coherent optical coupling of SAW cavity devices with frequencies of several GHz by changing the optical angle of incidence (see section S10 of supplementary information). The optomechanical coupling rate of devices can also be improved significantly through reduced acoustic mode volumes in cavities with smaller acoustic waists (see section S10 of supplementary information). Moreover, because the acoustic mode volume of the Gaussian SAW cavities scale inversely with the acoustic frequency, GHz SAW cavities naturally offer increased coupling strengths. Acoustic cavity losses can also be further improved by adopting etched groove reflectors in favor of metallic strips to eliminate both ohmic losses within the reflectors on piezoelectric substrates as well as additional acoustic losses within the reflectors.

A natural extension of the technique presented here would be to enclose the system within optical cavities. A SAW-mediated cavity optomechanical systems with an operation frequency of $\sim 4\ GHz$, Q-factors well exceeding $10^5$, and coupling rates comparable to nanomechanical systems($\frac{g_0}{2\pi} \sim 10$ kHz) could be achieved through straightforward improvements detailed in supplementary information section S10. The power-handling capability of this system, limited only by material damage, allows for large intracavity photon numbers ($n_c > 10^9$) which can consequently enable large optomechanical cooperativities ($C_{\mathrm{om}} > 1000$) (see section S10 of supplementary information). This platform therefore yields the high-power handling capability of bulk optomechanical systems[59,66] while also offering large coupling rates, small sizes, and ideal integrability to quantum systems and sensing devices.

A SAW cavity-optomechanical platform may have several straightforward applications. Strain fields of surface acoustic phonon modes can be readily coupled to a range of qubit systems, including spin qubits, quantum dots, and superconducting qubits, enabling novel quantum transduction strategies. Optical coupling to several other strain-sensitive quantum systems, including superfluids and 2D materials, can also be realized, which could yield new fundamental insights into novel condensed matter phenomena. The SAW-based cavity optomechanical system could also serve as an alternate platform for microwave-to-optical transduction schemes circumventing conventional challenges such as poor phonon-injection efficiencies, low-power handling capabilities, and fabrication challenges[7,49,50]. Beyond novel devices for quantum systems, the demonstrated techniques and devices also represent an attractive strategy for realizing a new class of non-contact all-optical SAW-based sensors with targets ranging from small molecules to large biological entities including viruses and bacteria, without electrical contacts or constraints. Moreover, in contrast to prior electromechanical techniques, the material versatility available to the optomechanical coupling presented enables broadly applicable material spectroscopy for basic studies of phonons and material science.

In summary, here we demonstrate coherent optical coupling to surface acoustic cavities on crystalline substrates. A novel non-collinear Brillouin-like parametric interaction accesses high-frequency Gaussian SAW cavity modes without the need for piezo-electric coupling, enabling record cavity quality factors. Optomechanical coupling in SAW cavities could be enabling for hybrid quantum systems, condensed matter physics, SAW-based sensing, and material spectroscopy. For hybrid quantum systems, this interaction, in conjunction with demonstrated techniques of strong coupling of SAWs to quantum systems (e.g. qubits, 2D materials, and superfluids), could form the basis for the next generation of hybrid quantum platforms. For sensing, this platform could enable a new class of SAW sensors agnostic to piezoelectric properties and free of electrical constraints and resulting parasitic effects. Finally, the coherent coupling technique enables detailed phonon spectroscopy of intrinsic mechanical loss mechanisms for a wide array of materials without the limitations of extrinsic probing devices.

## Methods

**Device Fabrication:** To fabricate the GaAs devices, a single crystal [100]-cut GaAs is coated with a PMMA polymer layer and the required reflector profiles are drawn on the polymer with an e-beam lithography tool. Subsequently, the required thickness of metal, in this case 200nm Aluminum, is deposited using an ultra-high vacuum e-beam evaporation tool system. Finally, the excess polymer is removed using an acetone bath to obtain the experimental devices. A more detailed description of the device fabrication is provided in section S4 of supplementary information.

**Numerical Methods:** Determining the exact acoustic reflector profiles requires the SAW group velocity as a function of angle from the chosen SAW cavity axis, i.e., the anisotropy of the substrate. This is calculated by numerically solving acoustic wave equations with appropriate boundary conditions. To efficiently confine SAW fields, the shape of the reflector must match the radius of curvature of the confined gaussian mode. The calculated group velocity can then be used to determine the radius of curvature of the reflectors as a function of the axial location and angle from the cavity axis ($R(x,\theta)$). These reflector profiles are imported into finite element software to validate the cavity designs by verifying the stability of high-Q Gaussian-like SAW modes (Fig. 1d-1f). A detailed description of the FEM simulation procedure is provided in section S3 of supplementary information.

**Phonon Spectroscopy:** A sensitive phonon-mediated four-wave mixing measurement technique is developed, building off of related techniques for measuring conventional Brillouin interactions. The SAW cavity mode is driven with two optical tones, which are incident at angles designed to target specific phonon frequencies. A probe beam at a disparate wavelength incident collinear to one of the drive tones scatters off the optically driven SAW cavity mode to generate the measured response. The angle of incidence of the optical fields is controlled through off-axis incidence on a well-calibrated aspheric focusing lens (see section S6 of supplementary information). The optomechanically scattered signal is collected on a single-mode collimator and spectrally filtered using a fiber-Bragg grating to reject excess drive light. The resulting signal is combined with a local oscillator (LO) and measured with a balanced detector (see section S5 of supplementary information). The measured signal is a coherent sum of frequency-independent Kerr four-wave mixing in the bulk of the crystalline substrate and the optomechanical response, giving rise to Fano-like resonances. This spectroscopy technique can resolve optomechanical responses with < fW optical powers. A detailed description of the experimental apparatus and the angle tuning technique is provided in sections S5 and S6 of supplementary information, respectively.

# Supplementary Information for:

# Coherent Optical Coupling to Surface Acoustic Wave Devices

Arjun Iyer[1], Yadav P. Kandel[2], Wendao Xu[1], John M. Nichol[2] and William H. Renninger[1,2]

[1] Institute of Optics, University of Rochester, Rochester, NY 14627, USA. [2]Department of Physics and Astronomy, University of Rochester, Rochester, NY 14627, USA. *email: aiyer2@ur.rochester.edu

# S1. Theoretical Estimation of Optomechanical Coupling Strength

In this section, we derive an analytic expression for the coupling rate for parametric coupling between traveling-wave non-collinear optical fields and a standing wave SAW cavity mode. The assumptions made during the derivation are minimal and provide a useful alternative to computationally intensive FEM calculations. A simple analytical expression allows the extraction of dependencies between coupling rate and material parameters.

The system is modeled as follows (Fig. S1)- a semi-infinite crystalline medium occupies the region $z < 0$ and supports surface acoustic waves confined to the material interface, $z = 0$. Periodic acoustic reflectors along the x-axis confine surface acoustic fields, resulting in a Gaussian SAW cavity. Two non-collinear optical waves, pump and stokes fields, which subtend equal but opposite angles ($\theta$) with the z-axis, are incident from outside the medium ($z > 0$) in the region enclosed by the two acoustic mirrors. Because optomechanical scattering is a vectorial process, the resulting optomechanical coupling is a function of the polarization of incident optical fields. We derive optomechanical coupling strengths for the following cases -1) both optical fields are TE polarized, 2) the pump is TE-polarized, and the Stokes field is TM-polarized (TE-TM), and 3) both fields are TM-polarized (TM-TM).

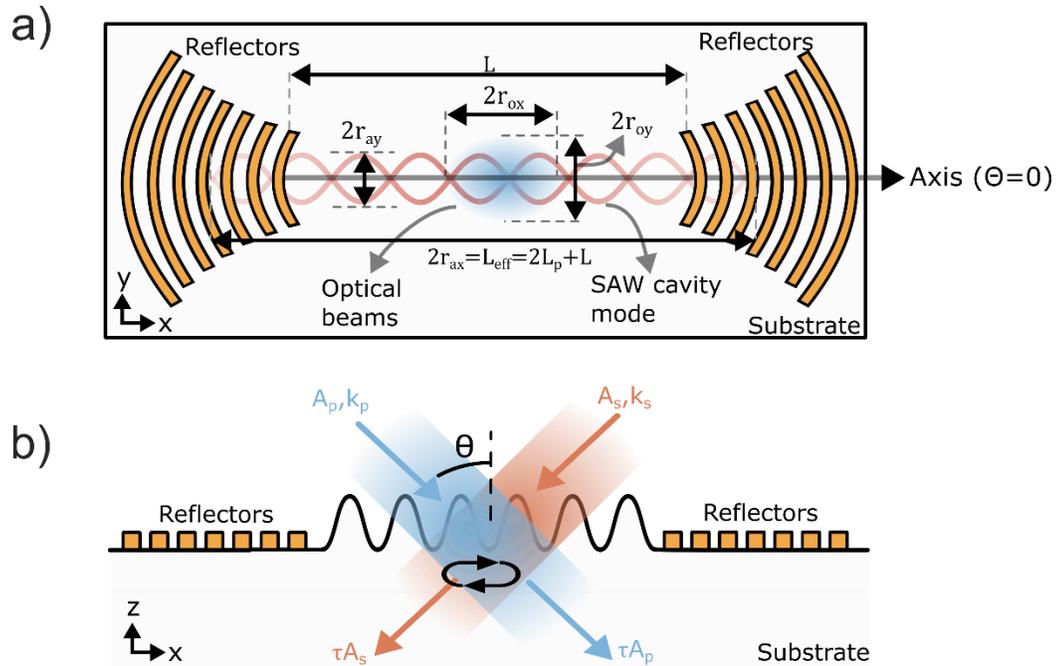

Figure S1: Theoretical estimation of optomechanical coupling rate of a SAW based optomechanical device. a) Top view (XY) of the optomechanical systems with the relevant optical and acoustic dimensions. b) Side view of the optomechanical device

### A) TE-polarized optical fields

In the case of TE-polarized optical fields, the electric field for the pump ($E_p$) and Stokes ($E_s$) fields outside the medium close to the surface ($z \rightarrow 0^+$) are given as-

$$E_p(x, y, z = 0^+) = A_p \exp\left(-\frac{x^2}{r_{ox}^2} - \frac{y^2}{r_{oy}^2}\right) \exp(ik_p \sin\theta\, x)\, \hat{y} \tag{1}$$

$$E_s(x, y, z = 0^+) = A_s \exp\left(-\frac{x^2}{r_{ox}^2} - \frac{y^2}{r_{oy}^2}\right) \exp(-ik_s \sin\theta\, x)\, \hat{y} \tag{2}$$

Where $A_p (A_s)$ and $k_p(k_s)$ refer to the amplitude and wavevector of the pump (Stokes) field, respectively. Since the acoustic frequencies are much smaller than optical frequencies, we assume $k_p \approx k_s = k_0$. $r_{ox}$ and $r_{oy}$ refer to the effective optical beam radius along the x and y-axis, respectively. Since the fields are incident at an angle, the resultant distribution on the surface is not symmetric along the x and y-axes. The effective beam waist along the x-axis ($r_{ox}$) can be expressed as $r_{ox} = r_0/\cos\theta$, while the beam waist along the y-axis remains unchanged, $r_{oy} = r_0$, where $r_0$ is the incident beam radius. The electric fields on the other side ($z \rightarrow 0^-$) of the interface can be readily derived from Eqn.1 and Eqn. 2 by multiplying the appropriate Fresnel transmission coefficients ($\tau(\theta)$) as follows.

$$E_p(x, y, z = 0^-) = A_p \tau(\theta) \exp\left(-\frac{x^2}{r_{ox}^2} - \frac{y^2}{r_{oy}^2}\right) \exp(ik_0 \sin\theta\, x)\, \hat{y} \tag{3}$$

$$E_s(x, y, z = 0^-) = A_s \tau(\theta) \exp\left(-\frac{x^2}{r_{ox}^2} - \frac{y^2}{r_{oy}^2}\right) \exp(-ik_0 \sin\theta\, x)\, \hat{y} \tag{4}$$

Accounting for the Gaussian mode profile of the $m^{th}$ SAW cavity mode, the acoustic field can be expressed as[1,2]

$$u_x = U_0(\exp(-\eta q_m z - i\phi) + c.c.) \exp\left(-\frac{x^2}{r_{ax}^2} - \frac{y^2}{r_{ay}^2}\right) \cos(q_m x) \tag{5}$$

$$u_z = -\frac{U_0}{i}(\gamma\exp(-\eta q_m z - i\phi) + c.c.) \exp\left(-\frac{x^2}{r_{ax}^2} - \frac{y^2}{r_{ay}^2}\right) \cos(q_m x) \tag{6}$$

Where, $u_x, u_z$ are the x and z-component of the acoustic displacement, respectively. $U_0, q_m, r_{ax}, r_{ay}$ refer to the amplitude, wavevector of the $m^{th}$ cavity mode, waist of the cavity mode along the x-axis given by $r_{ax} = L_{eff}/2$, and waist along the y-axis, respectively. $\eta, \phi, \gamma$ are material-dependent parameters obtained by solving the acoustic wave equation with appropriate boundary conditions[1,2]. 'c.c' refers to the complex conjugate of the term preceding it.

Given the acoustic and electric fields, we define the traveling wave coupling rate ($g_0$) based off conventional definitions as[3]-

$$g_0 = -\frac{\omega x_{zpf}}{2} \frac{\langle f \cdot u \rangle}{N_p N_s} \qquad (7)$$

Where $\omega, f, u$ refer to the optical frequency, optical force distribution, and acoustic field distribution. The acousto-optic overlap is defined through an overlap integral given by- $\langle f \cdot u \rangle = \int f \cdot u^* dV$. $x_{zpf}$ is the zero-point displacement of the mechanical mode defined as $x_{zpf} = \sqrt{\frac{\hbar}{2m_{eff}\Omega_m}}$, $\Omega_m$ refers to the frequency of the acoustic mode, $m_{eff} = \int \rho |u|^2 dV$ is the effective mass of the mechanical mode, and $\rho$ is the density of the substrate. $N_{p(s)} = \sqrt{\frac{1}{2}\epsilon_0 \int \epsilon |E_{p(s)}|^2 dV}$ serves as the power normalization factor. The volume integrals are performed over the effective interaction volume, which in the case of SAWs, is a acoustic wavelength-sized slice of material near the interface, within which energy of the SAWs are confined to.

Note that the traveling-wave coupling rate, as defined in this work, is similar in form to the coupling rate defined in conventional cavity optomechanics[3,4], except that the integrals are only performed over the interaction volume and not over the entire optical cavity, as would be the case in standard cavity-optomechanical calculations[4,5]. An equivalent cavity optomechanical ($g_0^c$) coupling rate can be derived from $g_0$ as follows-

$$g_0^c = g_0 \left(\frac{l_a}{l_{opt}}\right) \qquad (8)$$

Where $l_a$ is the effective length of the interaction volume as used when determining $g_0$, and $l_{opt}$ is the length of the optical cavity. Alternatively, the traveling-wave coupling rate, $g_0$, can be understood as being the largest possible cavity optomechanical coupling rate possible in an SAW-based cavity optomechanical system, achieved when the optical cavity mode and acoustic mode perfectly overlap, i.e. $l_{opt} = l_a$.

The strength of interactions studied in this work can also be quantified through Brillouin-like Gain coefficient (G) defined as[6]-

$$G = \frac{\omega Q_m}{2\Omega_m^2 P_p P_s} \frac{|\langle f \cdot u^* \rangle|^2}{\langle u \cdot \rho u \rangle} \qquad (9)$$

Where $P_p, P_s$ are incident pump and Stokes powers, and $Q_m$ is the mechanical quality factor. The coupling rate and the Brillouin gain coefficient are related as

$$|g_0| = \left(\frac{v_p v_s \hbar \omega_p \Omega_m G}{4 l_a^2 Q_m}\right)^{1/2} \qquad (10)$$

Next, we derive the acousto-optic contributions from optical forces, namely, radiation pressure and electrostriction on the surface and the bulk of the medium.

**Radiation Pressure**

Radiation pressure force denoted as $P_{rp}$ is given by[7]

$$P_{rp} = \frac{1}{2}\epsilon_0 \; E_{pt} E_{st}^*(\epsilon - 1) - \frac{1}{2}\epsilon_0^{-1} D_{pn} D_{sn}^* \left(\frac{1}{\epsilon} - 1\right) \qquad (11)$$

Where $E_{p(s)t(n)}, D_{p(s)t(n)}$ refer to tangential (normal) pump (Stokes) electric and displacement fields. $\epsilon_0, \epsilon$ refer to the dielectric permittivities of the vacuum and the material, respectively. Radiation pressure force points along the surface normal, i.e., along the positive z-axis. For TE fields $E_{pn} = E_{sn} = 0$, and the resultant expression for $P_{rp}$ can be simplified as follows-

$$P_{rp}(x,y) = \frac{1}{2}\epsilon_0(\epsilon - 1)|\tau|^2 \; A_p A_s^* e^{-2x^2/r_{0x}^2} e^{-2y^2/r_0^2} e^{i2k_0 \sin\theta \; x} \; \hat{z} \qquad (12)$$

The resulting radiation pressure force would only overlap with the z-component of the acoustic displacement. The acoustic-optic overlap can be expressed as

$$\langle f \cdot u \rangle_{rp} = \int_{-\infty}^{\infty} dx \, dy \; P_{rp}(x,y) u_z^*(z=0,x,y) \qquad (13)$$

$$= \epsilon_0 \frac{\epsilon - 1}{2}|\tau|^2 \; A_p A_s^* \; \frac{U_0}{i} \; 2\text{Re}(e^{-i\phi}\gamma) \int_{-\infty}^{\infty} dx \; e^{-ax^2} e^{i\Delta kx} \cos q_m x \; \int_{-\infty}^{\infty} dy \; e^{-by^2} \qquad (14)$$

Where we define additional parameters $a, \Delta k$ and $b$ as

$$a = \frac{2}{r_{ox}^2} + \frac{1}{r_{ax}^2}, \qquad (15)$$

$$b = \frac{2}{r_0^2} + \frac{1}{r_{ay}^2} \qquad (16)$$

$$\Delta k = 2k_0 \sin\theta. \qquad (17)$$

Since the integrals concerning the spatial variables x and y are independent, we separately calculate the two integrals

$$I_1 = \int_{-\infty}^{\infty} dy\, e^{-by^2} = \sqrt{\frac{\pi}{b}} \qquad (18)$$

$$I_2 = \int_{-\infty}^{\infty} dx\, e^{-ax^2} e^{i\Delta kx} \cos q_m x = \int_{-\infty}^{\infty} dx\, e^{-ax^2} e^{i\Delta kx} \frac{1}{2}(e^{-iq_m x} + e^{+iq_m x}) = \frac{1}{2} e^{-\frac{(\Delta q)^2}{4a}} \sqrt{\frac{\pi}{a}} \qquad (19)$$

Where $\Delta q = \Delta k - q_m$ is the wavevector difference between the optical forces and the SAW cavity mode. While deriving Eqn. 19, the term which corresponds to the part of the standing-wave acoustic mode not phase matched to the optical forces ($\exp(i(\Delta k + q_m)x)$) is neglected, similar to rotating wave approximation in atomic physics. Note in Eqn. 19 that the resulting acousto-optic overlap is a function of phase-mismatch $\Delta q$ and is maximum when the optical forces are perfectly phase matched with the acoustic cavity mode ($\Delta q = 0$, i.e., $\Delta k = q_m$). This dependence results in the phase matching effects outlined in the main text. The next section will explore further implications of acoustic overlap and coupling rate dependence on phase-mismatch. For the subsequent calculations in this section, we assume $\Delta q = 0$.

The acousto-optic overlap resulting from radiation pressure forces can now be expressed as-

$$\langle f \cdot u \rangle_{rp} = \epsilon_0 \frac{\epsilon-1}{2i} |\tau|^2 U_0 A_p A_s^* \operatorname{Re}(e^{-i\phi}\gamma) \frac{\pi}{\sqrt{ab}} \qquad (20)$$

**Photoelastic forces**

Time-varying electric fields within a dielectric material can generate time-varying photoelastic optical forces. Photoelastic stresses resulting in optical forces are derived from the photoelastic tensor. For a material with a cubic crystalline lattice whose principal axes are oriented along the assumed cartesian axis, the stress tensor in the Voigt notation is now given[7,8]-

$$\begin{pmatrix} \sigma_{xx} \\ \sigma_{yy} \\ \sigma_{zz} \\ \sigma_{zy} \\ \sigma_{zx} \\ \sigma_{xy} \end{pmatrix} = -\frac{1}{2}\epsilon_0 n^4 \begin{pmatrix} p_{11} & p_{12} & p_{12} & 0 & 0 & 0 \\ p_{12} & p_{11} & p_{12} & 0 & 0 & 0 \\ p_{12} & p_{12} & p_{11} & 0 & 0 & 0 \\ 0 & 0 & 0 & p_{44} & 0 & 0 \\ 0 & 0 & 0 & 0 & p_{44} & 0 \\ 0 & 0 & 0 & 0 & 0 & p_{44} \end{pmatrix} \begin{pmatrix} E_{px}E_{sx}^* \\ E_{py}E_{sy}^* \\ E_{pz}E_{sz}^* \\ E_{pz}E_{sy}^* + E_{py}E_{sz}^* \\ E_{pz}E_{sx}^* + E_{px}E_{sz}^* \\ E_{py}E_{sx}^* + E_{px}E_{sy}^* \end{pmatrix} \quad (21)$$

Here we have invoked the crystal symmetry to assume $p_{12} = p_{13} = p_{32}$, this may not be true if the coordinate system does not coincide with the principal crystal axes. The pump and Stokes electric fields are calculated within the material. For TE fields, $E_{px} = E_{sx} = E_{pz} = E_{sz} = 0$ and the resultant stresses are-

$$\sigma_{xx} = -\frac{1}{2}\epsilon_0 n^4 p_{12} E_{py} E_{sy}^* \quad (22)$$

$$\sigma_{yy} = -\frac{1}{2}\epsilon_0 n^4 p_{11} E_{py} E_{sy}^* \quad (23)$$

$$\sigma_{zz} = -\frac{1}{2}\epsilon_0 n^4 p_{12} E_{py} E_{sy}^* \quad (24)$$

$$\sigma_{xy} = \sigma_{yz} = \sigma_{zx} = 0 \quad (25)$$

In a system comprising homogeneous materials, photoelastic forces can exist inside each material, resulting in body forces in the bulk of the medium and at material interfaces where discontinuous stresses are present, resulting in surface pressure (analogous to radiation pressure). We separately calculate the contribution to the acousto-optic overlap of photoelastic forces on the surface and within the bulk of the medium.

The excess photoelastic surface force on the interface ($z = 0$) is given as

$$P_{es} = \sigma_{xz}(z=0)\,\hat{x} + \sigma_{yz}(z=0)\hat{y} + \sigma_{zz}(z=0)\hat{z} = -\frac{1}{2}\epsilon_0 n^4 p_{12} E_{py} E_{sy}^* \hat{z} \quad (26)$$

The acosuto-optic overlap resulting from photoelastic surface pressure is given by-

$$\langle f \cdot u \rangle_{es} = \int_{-\infty}^{\infty} dx\, dy\, P_{ess}(x,y) U_z^*(z=0,x,y) \qquad (27)$$

$$\langle f \cdot u \rangle_{es} = -\frac{1}{2i} n^4 p_{12} \epsilon_0 |\tau|^2\, U_0 A_p A_s^*\, \text{Re}(e^{-i\phi}\gamma) \frac{\pi}{\sqrt{ab}} \qquad (28)$$

Next, the overlap resulting from bulk photoelastic forces is calculated. The photoelastic body forces can be determined from the divergence of stress components and its vectorial components are given as[8,9]-

$$f_x = -\partial_x \sigma_{xx} - \partial_y \sigma_{xy} - \partial_z \sigma_{xz} = -i\Delta k \sigma_{xx} - \partial_x \sigma_{xx} \qquad (29)$$

$$f_y = -\partial_x \sigma_{xy} - \partial_y \sigma_{yy} - \partial_z \sigma_{yz} = -\partial_y \sigma_{yy} \qquad (30)$$

$$f_z = -\partial_x \sigma_{xz} - \partial_y \sigma_{zy} - \partial_z \sigma_{zz} = -\partial_z \sigma_{zz} \propto \partial_z(E_p E_s) \qquad (31)$$

Since the SAW cavity modes of interest have no displacement along the y-axis, the only forces of concern are $f_x$ and $f_z$ along which acoustic displacements are non-zero. Note that the forces along the z-direction result from an electric field gradient along the z-axis. For beam sizes used in the experiments (~30 µm @ 1550 nm), which have typical optical Rayleigh lengths of $x_R^o \sim 1$ mm, the optical fields diffractive minimally along the z-axis, within the decay length of the SAW cavity mode (~5 µm), and therefore $f_z \sim 0$. Only the forces along the x-axis result in non-zero overlap with the SAW cavity mode.

Further expanding Eqn. 31, we get-

$$f_x = -\frac{1}{2}\epsilon_0\, \epsilon^2\, p_{12} A_p A_{s0}^* |\tau|^2 e^{-\frac{2x^2}{r_{ox}^2}} e^{-\frac{2y^2}{r_0^2}} e^{i2k_0 \sin\theta\, x} \left(-i\Delta k + \frac{4x}{r_{ox}}\right) \qquad (32)$$

The optomechanical overlap contribution from the bulk electrostricitve forces is now given by-

$$\langle f \cdot u \rangle_{eb} = -\frac{1}{2}\epsilon_0\, \epsilon^2\, p_{12} A_p A_{s0}^* |\tau|^2 U_0 \int_{-\infty}^{\infty} dy\, e^{-by^2} \int_{-\infty}^{0} dz\, (\exp(\eta q z - i\phi) +$$
$$\text{c.c.})^* \int_{-\infty}^{\infty} dx\, \frac{1}{2}\left(-iq_m + \frac{4x}{r_{ox}}\right) e^{i\Delta qx - ax^2} \qquad (33)$$

Eqn. 35 can be further simplified to-

$$\langle f \cdot u \rangle_{eb} = -\frac{1}{2i}\epsilon_0\, \epsilon^2\, p_{12} A_p A_{s0}^* |\tau|^2 U_0 \text{Re}\left(\frac{e^{-i\phi}}{\eta}\right) \frac{\pi}{\sqrt{ab}} \qquad (34)$$

The total optomechanical overlap can now be written as-

$$\langle f_{tot} \cdot u^* \rangle = \langle f \cdot u \rangle_{rp} + \langle f \cdot u \rangle_{es} + \langle f \cdot u \rangle_{eb} = \frac{i}{2}\epsilon_0 A_p A_s^* |\tau|^2 U_0 \frac{\pi}{\sqrt{ab}} \alpha \quad (35)$$

Here we define $\alpha$ as-

$$\alpha = \left( -(\epsilon - 1)\text{Re}(e^{-i\phi}\gamma) + p_{12}\epsilon^2 \text{Re}(e^{-i\phi}\gamma) + p_{12}\epsilon^2 \text{Re}\left(\frac{e^{-i\phi}}{\eta}\right) \right) \quad (36)$$

Next we calculate the effective mass of the acoustic mode given by-

$$m_{eff} = \langle u, \rho u \rangle = \int dV\, \rho\, (|u_x|^2 + |u_z|^2) \quad (37)$$

Inserting expressions for acoustic displacements from Eqn. 5 and Eqn.6 into Eq.39 and performing the requisite integrals one obtains-

$$\langle u, \rho u \rangle_V = |U_0|^2 \frac{\pi}{4q} r_{ay} r_{ax} \left( \frac{1}{\text{Re}(\eta)} + \text{Re}\left(\frac{e^{-2i\phi}}{\eta}\right) + \frac{|\gamma|^2}{\text{Re}(\eta)} + \text{Re}\left(\frac{\gamma^2}{\eta}e^{-i2\phi}\right) \right) \quad (38)$$

We rewrite Eqn. 38 compactly as

$$\langle u, \rho u \rangle_V = |U_0|^2 \rho \frac{\pi}{4q} r_{ay} r_{ax} \delta \quad (39)$$

Where we define $\delta$ as

$$\delta = \frac{1}{\text{Re}(\eta)} + \text{Re}\left(\frac{e^{-2i\phi}}{\eta}\right) + \frac{|\gamma|^2}{\text{Re}(\eta)} + \text{Re}\left(\frac{\gamma^2}{\eta}e^{-i2\phi}\right) \quad (40)$$

Next, we calculate the pump and Stokes power normalization factors $N_p$ and $N_s$-

$$N_{p(s)} = \sqrt{\frac{1}{2}\epsilon_0 \int \epsilon |E_{p(s)}|^2 dV} = \left(\frac{1}{2}\epsilon_0 \epsilon |A_{p(s)}|^2 |\tau|^2 \left(\frac{\pi r_0^2}{2}\right) l_a \right)^{\frac{1}{2}} \quad (41)$$

Where $l_a$ is the acoustic decay length defined as $l_a = (2\text{Re}(\eta)q)^{-1}$. The acoustic length can be understood as the characteristic length within which the energy of the surface acoustic field decays within the bulk of the substrate.

Combining Eqn. 7, 35,39 and 41 the coupling rate can be expressed as-

$$g_0 = \frac{-i\omega\alpha}{\epsilon r_0^2 l_a} \sqrt{\frac{2\hbar q_m}{r_{ay} r_{ax} \delta ab \rho \Omega_m \pi}} \quad (42)$$

Equivalently, the Brillouin gain coefficient can be given as-

$$G_{TE} = \frac{8q\omega Q}{\Omega_0^2 c^2 \rho \pi} \times \frac{|\alpha|^2}{\epsilon\delta} \times \frac{1}{ab\, r_{ay} r_{ax} r_0^4} \quad (43)$$

### B) TE-polarized pump and TM-polarized Stokes optical fields

For the case of cross-polarized optical fields, the pump field is assumed to be TE-polarized, as in Case A, and the Stokes field is TM-polarized. The electric fields inside the medium, close to the surface $(z \to 0^-)$ are given as-

$$E_p(x, y, z = 0^-) = A_p \tau \exp\left(-\frac{x^2}{r_{0x}^2} - \frac{y^2}{r_0^2}\right) \exp(ik_0 \sin\theta\, x)\, \hat{y} \quad (44)$$

$$E_s(x, y, z = 0^+) = A_s \tau (-\cos\theta\, \hat{x} + \sin\theta\, \hat{z}) \exp\left(-\frac{x^2}{r_{0x}^2} - \frac{y^2}{r_0^2}\right) \exp(-ik_0 \sin\theta\, x) \quad (45)$$

**Radiation Pressure**

Since the pump and Stokes optical fields do not have overlapping non-zero electric field components (since they are perpendicularly polarized), the net radiation pressure is zero, and by extension, the acoustic overlap is zero.

$$\langle f \cdot u \rangle_{rp} = 0 \quad (46)$$

**Photoelastic forces**

Photoelastic stresses for cross-polarized optical fields, as in Case A is, given by-

$$\begin{pmatrix}\sigma_{xx}\\ \sigma_{yy}\\ \sigma_{zz}\\ \sigma_{zy}\\ \sigma_{zx}\\ \sigma_{xy}\end{pmatrix} = -\frac{1}{2}\epsilon_0 n^4 \begin{pmatrix} p_{11} & p_{12} & p_{12} & 0 & 0 & 0\\ p_{12} & p_{11} & p_{12} & 0 & 0 & 0\\ p_{12} & p_{12} & p_{11} & 0 & 0 & 0\\ 0 & 0 & 0 & p_{44} & 0 & 0\\ 0 & 0 & 0 & 0 & p_{44} & 0\\ 0 & 0 & 0 & 0 & 0 & p_{44}\end{pmatrix}\begin{pmatrix}E_{px}E_{sx}^*\\ E_{py}E_{sy}^*\\ E_{pz}E_{sz}^*\\ E_{pz}E_{sy}^*+E_{py}E_{sz}^*\\ E_{pz}E_{sx}^*+E_{px}E_{sz}^*\\ E_{py}E_{sx}^*+E_{px}E_{sy}^*\end{pmatrix} \quad (47)$$

Here Fresnel transmission coefficient for TE and TM polarization is assumed to be approximately equal. The resultant photoelastic stresses are given by-

$$\sigma_{xx} = \sigma_{yy} = \sigma_{zz} = \sigma_{zx} = 0 \quad (48)$$

$$\sigma_{zy} = -\frac{1}{2}\epsilon_0 n^4 p_{44} E_{py} E_{sz}^* \quad (49)$$

$$\sigma_{xy} = -\frac{1}{2}\epsilon_0 n^4 p_{44} E_{py} E_{sx}^* \quad (50)$$

Analogous to Eqn. 28, the photoelastic surface force is now given by-

$$P_{ess} = \frac{1}{2}\epsilon_0 n^4 p_{44} E_{py} E_{sz}^* \, \hat{y} \quad (51)$$

Since SAW cavity modes of interest have no displacement component along the y-axis, the resultant acoustic-optic overlap resulting from electrostrictive surface forces is then

$$\langle f \cdot u \rangle_{es} = 0 \quad (52)$$

Vectorial components of the photoelastic body forces are determined as in Eqn. 31, Eqn. 32, and Eqn. 33 as

$$f_x = -iq\,\partial_x\sigma_{xx} - \partial_y\sigma_{xy} - \partial_z\sigma_{xz} = -\partial_y\sigma_{xy} \quad (53)$$

$$f_y = -iq\,\partial_x\sigma_{xy} - \partial_y\sigma_{yy} - \partial_z\sigma_{yz} = -iq_m\,\partial_x\sigma_{xy} \quad (54)$$

$$f_z = -iq\,\partial_x\sigma_{xz} - \partial_y\sigma_{zy} - \partial_z\sigma_{zz} = -\partial_y\sigma_{zy} \quad (55)$$

As before, only the x and z-component of the photoelastic forces have non-zero overlap with acoustic displacements since the SAW cavity mode does not displacements along the y-axis. The forces along x and z-axis can be further expanded as-

$$f_x = \frac{1}{2}\epsilon_0 \, \epsilon^2 \, p_{44} A_p A_{s0}^* |\tau|^2 \cos\theta \left(-\frac{4y}{r_0^2}\right) e^{-\frac{2x^2}{r_{0x}^2}} e^{-\frac{2y^2}{r_0^2}} \quad (56)$$

$$f_z = \frac{1}{2}\epsilon_0 \, \epsilon^2 p_{44} A_p A_{s0}^* |\tau|^2 \sin\theta \left(-\frac{4y}{r_0^2}\right) e^{-\frac{2x^2}{r_{0x}^2}} e^{-\frac{2y^2}{r_0^2}} \quad (57)$$

The acousto-optic overlap corresponding to forces along x- and z-axis can now be determined as-

$$\langle f_x \cdot u_x^* \rangle = \frac{1}{2}\epsilon_0 \, \epsilon^2 \, p_{44} A_p A_{s0}^* |\tau|^2 \cos\theta \, U_0 \int_{-\infty}^{\infty} dy \left(\frac{4y}{r_0^2}\right) e^{-by^2} \int_0^{\infty} dz \, (\exp(-\eta q z - i\phi) + \text{H.C.})^* \int_{-\infty}^{\infty} dx \, e^{-ax^2} \quad (58)$$

Since $\int_{-\infty}^{\infty} dy \left(\frac{4y}{r_0^2}\right) e^{-by^2} = 0$, the contribution resulting from the x-component of photoealstic forces is $\langle f_x \cdot u_x \rangle = 0$

Similarly, the z-component of the photoelastic body force also yields no overlap $\langle f_z \cdot u_z \rangle = 0$

As a consequence, the total bulk electrostriction overlap is-

$$\langle f \cdot u \rangle_{eb} = 0 \quad (59)$$

As a result, the total overlap and, consequently, the optomechanical coupling rate for this configuration is

$$g_0^{TE-TM} = 0 \quad (60)$$

The absence of optomechanical coupling for the TE-TM scattering is primarily a result of the assumed crystal symmetry (cubic) and the resulting symmetry in the photoelastic tensor. In crystal structures with reduced symmetry, such as crystalline quartz and LiNbO$_3$, SAW mediated optomechanical processes can couple orthogonal polarizations.

### C) TM-polarized optical fields

For the case where both pump and Stokes fields are TM-polarized, the electric fields inside the medium close to the surface ($z \to 0^-$) are given as-

$$E_p(x, y, z = 0^-) = A_p\tau(-\cos\theta\ \hat{x} - \sin\theta\ \hat{z})\exp\left(-\frac{x^2}{r_{0x}^2} - \frac{y^2}{r_0^2}\right)\exp(ik_0\sin\theta\ x) \quad (61)$$

$$E_s(x, y, z = 0^-) = A_s\tau(-\cos\theta\ \hat{x} + \sin\theta\ \hat{z})\exp\left(-\frac{x^2}{r_{0x}^2} - \frac{y^2}{r_0^2}\right)\exp(-ik_0\sin\theta\ x) \quad (62)$$

**Radiation Pressure**

Radiation pressure force, similar to cases A and B can be expressed as-

$$P_{rp} = \frac{1}{2}\epsilon_0(\epsilon - 1)A_p A_s^* |\tau|^2 \exp\left(-\frac{2x^2}{r_{0x}^2} - \frac{2y^2}{r_0^2}\right) e^{i2k_0 \sin\theta\ x}(\cos^2\theta - \epsilon\sin^2\theta)\ \hat{z} \quad (63)$$

The corresponding acousto-optic overlap is given as-

$$\langle f \cdot u \rangle_{rp} = \int_{-\infty}^{\infty} dx\ dy\ P_{rp}(x,y) U_z^*(z=0,x,y) \quad (64)$$

$$\langle f \cdot u \rangle_{rp} = \epsilon_0 \frac{(\epsilon-1)(\cos^2\theta - \epsilon\sin^2\theta)}{2i} |\tau|^2\ U_0 A_p A_s^*\ \text{Re}(e^{-i\phi}\gamma)\frac{\pi}{\sqrt{ab}} \quad (65)$$

**Photoelastic forces**

The components of the photoelastic stress tensor are-

$$\sigma_{xx} = -\frac{1}{2}\epsilon_0 n^4 p_{11} E_{px} E_{sx}^* - \frac{1}{2}\epsilon_0 n^4 p_{12} E_{pz} E_{sz}^* \quad (66)$$

$$\sigma_{yy} = 0 \quad (67)$$

$$\sigma_{zz} = -\frac{1}{2}\epsilon_0 n^4 p_{12} E_{px} E_{sx}^* - \frac{1}{2}\epsilon_0 n^4 p_{11} E_{pz} E_{sz}^* \quad (68)$$

$$\sigma_{xz} = -\frac{1}{2}\epsilon_0 n^4 p_{44}(E_{pz}E_{sx}^* + E_{px}E_{sz}^*) = 0 \quad (69)$$

$$\sigma_{xy} = \sigma_{yz} = 0 \quad (70)$$

The electrostriction surface forces on the interface is given by-

$$P_{ess} = -\frac{1}{2}\epsilon_0 n^4 A_p A_s^* |\tau|^2 (p_{12}\cos^2\theta - p_{11}\sin^2\theta)\ \hat{z} \quad (71)$$

The resulting overlap with acoustic mode is given as-

$$\langle f \cdot u \rangle_{es} = \int_{-\infty}^{\infty} dx\, dy\, P_{ess}(x,y) U_z^*(z=0,x,y)$$

$$\langle f \cdot u \rangle_{es} = \frac{-1}{2i} n^4 (p_{12} \cos^2\theta - p_{11} \sin^2\theta) \epsilon_0 |\tau|^2 U_0 A_p A_s^* \, \mathrm{Re}(e^{-i\phi}\gamma) \frac{\pi}{\sqrt{ab}} \qquad (72)$$

The electrostrictive body forces in the bulk of the substrate are given by

$$f_x = -iq\, \partial_x \sigma_{xx} - \partial_y \sigma_{xy} - \partial_z \sigma_{xz} = -iq_m\, \partial_x \sigma_{xx} \qquad (73)$$

$$f_y = -iq\, \partial_x \sigma_{xy} - \partial_y \sigma_{yy} - \partial_z \sigma_{yz} = 0 \qquad (74)$$

$$f_z = -iq\, \partial_x \sigma_{xz} - \partial_y \sigma_{zy} - \partial_z \sigma_{zz} = -\partial_z \sigma_{zz} \sim 0 \qquad (75)$$

The z-component of the body force, $f_z$ is assumed to be zero following reasoning from case A (TE-TE scattering). The resulting overlap is expressed as-

$$\langle f \cdot u \rangle_{es} = \int_V dV\, f_x u_x + f_y u_y + f_z u_z = \int_V dV\, f_x u_x$$

$$= \frac{i}{2} \epsilon_0 \epsilon^2 (p_{11} \cos^2\theta - p_{12} \sin^2\theta) A_p A_{s0}^* |\tau|^2 U_0 \mathrm{Re}\left(\frac{e^{i\phi}}{\eta}\right) \frac{\pi}{\sqrt{ab}} \qquad (76)$$

The total overlap is then given as-

$$\langle f_{tot} \cdot u^* \rangle = \frac{i}{2} \epsilon_0 A_p A_{s0}^* |\tau|^2 U_0 \frac{\pi}{\sqrt{ab}} \Big( (\epsilon - 1)(\cos^2\theta - \epsilon \sin^2\theta) \mathrm{Re}(e^{-i\phi}\gamma) - (p_{12} \cos^2\theta -$$

$$p_{11} \sin^2\theta)\, \epsilon^2 \mathrm{Re}(e^{-i\phi}\gamma) - (p_{11} \cos^2\theta - p_{12} \sin^2\theta) \epsilon^2 \mathrm{Re}\left(\frac{e^{i\phi}}{\eta}\right) \Big) \qquad (77)$$

The optomechanical coupling rate for the TM-TM scattering process is then be given by-

$$g_0 = \frac{-i\omega \alpha_{TM}}{\epsilon r_0^2 l_a} \sqrt{\frac{2\hbar q_m}{r_{ay} r_{ax} \delta ab \rho \Omega_m \pi}} \qquad (78)$$

Where $\alpha_{TM}$ is defined as-

$$\alpha_{TM} = \left((\epsilon - 1)(\cos^2\theta - \epsilon \sin^2\theta)\text{Re}(e^{-i\phi}\gamma) - \epsilon^2(p_{12}\cos^2\theta - \right.$$

$$\left. p_{11}\sin^2\theta)\text{Re}(e^{-i\phi}\gamma) - \epsilon^2(p_{11}\cos^2\theta - \sin^2\theta)\text{Re}\left(\frac{e^{i\phi}}{\eta}\right)\right) \quad (79)$$

The corresponding Brillouin Gain coefficient is given by-

$$G_{TM} = \frac{8q\omega Q}{\Omega_0^2 c^2 \rho \pi} \times \frac{|\alpha_{TM}|^2}{\epsilon\delta} \times \frac{1}{ab\, r_{ay} r_{ax} r_0^4} \quad (80)$$

Note that strength of the TM-TM scattering process strongly depends on the optical angle of incidence and at larger angles can be significantly stronger than TE-TE scattering process.

The following parameter material values are used for GaAs to calculate coupling rates quoted in the main text.

| Parameter | Value | Description |
|---|---|---|
| $\lambda_o$ | 1550.05 nm | Optical wavelength |
| n | 3.37 | [100]- Cut GaAs refractive index |
| $r_0$ | 33 μm | Optical beam radius |
| $\omega/2\pi = c/\lambda_o$ | 193.55 THz | Optical frequency |
| $p_{11}$ | −0.165 | Photoelastic constant of GaAs |
| $p_{12}$ | −0.140 | Photoelastic constant of GaAs |
| $\rho$ | 5307 kg m$^{-3}$ | GaAs density |
| $\theta$ | 7.8° | The optical angle of incidence |
| $\theta_m = \theta$ | 7.8° | Phase-matched angle corresponding to SAW mode |
| $\lambda_a$ | 5.7 μm | Acoustic wavelength |
| L | 505 μm | Acoustic mirror separation |
| $L_p$ | $7\lambda_a$ | SAW penetration depth |
| $L_{eff} = L + 2L_p$ | 620 μm | Effective SAW cavity length |
| $r_{ax} = L/2$ | 292.6 μm | The effective radius of SAW mode along the x-axis |
| $r_{0a}$ | $4\lambda_a = 24$ μm | Acoustic waist radius along the y-axis |

## [110]-oriented cavities

The acoustic parameters characterizing SAWs along [110]-direction on [100]-cut GaAs are[1]

| Parameter | Value | Description |
|---|---|---|
| $v_R$ | 2865 m/s | Rayleigh SAW velocity |
| $\eta$ | $0.5 + 0.48i$ | SAW decay parameter |
| $\gamma$ | $0.68 - 1.16i$ | Parameter quantifying the ratio of SAW displacements along the x and z-axis |
| $\phi$ | 1.05 | Phase lag between x and z-components of acoustic displacement |
| Q | 7000 | Acoustic quality factor |

Since the SAW direction of propagation is not along a principal crystalline axis, the photoelasticity tensor will have to be rotationally transformed and as a result $p_{12} \neq p_{13}$. The transformed coefficients of interest are- $p_{12} = -0.078$ and $p_{13} = -0.140$. The coupling rate and Brillouin gain coefficient for TE-TE scattering process within [110]-oriented cavities can be evaluated to be $\frac{g_0}{2\pi} = 1.8 \times 10^3$ Hz and $G = 9 \times 10^{-7}$ W$^{-1}$

## [110]-oriented cavities

The acoustic parameters characterizing SAWs along [110]-direction on [100]-cut GaAs are[2]

| Parameter | Value | Description |
|---|---|---|
| $v_R$ | 2615 m/s | Rayleigh SAW velocity |
| $\eta$ | $0.40 + 0.56i$ | SAW decay parameter |
| $\gamma$ | $0.37 - 1.1i$ | Parameter quantifying the ratio of SAW displacements along the x and z-axis |
| φ | 0.95 | Phase lag between x and z-components of acoustic displacement |
| Q | 120,000 | Acoustic quality factor |

The resultant TE-TE coupling rate and Brillouin gain coefficient for [100]-oriented cavities can be evaluated to be $\frac{g_0}{2\pi} = 1.73 \times 10^3$ Hz and $G = 2.3 \times 10^{-5}$ W$^{-1}$

## S2. Phase matching envelope

When calculating the acousto-optic overlaps in Section 2, as in Eqn. 19, optical fields were assumed to be perfectly phase-matched to acoustic modes $\Delta k = q_m$. In the absence of the above assumption, the dependence of the optomechanical coupling rate on phase mismatch can be expressed as-

$$g_0 \propto \langle f \cdot u \rangle \propto e^{-\frac{(\Delta q)^2}{4a}}, \qquad (81)$$

where $\Delta q = q_m - \Delta k$ quantifies the phase mismatch, and parameter $a$ is defined in Eqn. 15. The Gaussian dependence of coupling rate on phase mismatch is evident in Eqn. 81. The characteristic width is quantified by the parameter $\delta k = 2\sqrt{a} = 2\left(\frac{2}{r_{ox}^2} + \frac{1}{r_{ax}^2}\right)^{1/2}$. For experimental cavities investigated in this work, the acoustic cavity length ($r_{ax}$) is much larger than optical beam sizes, i.e. $r_x \ll r_{ax}$, and as a result, $\delta k = 2\sqrt{a} \approx \frac{2\sqrt{2}}{r_{ox}} = \frac{2\sqrt{2}\cos\theta}{r_0}$. For small $\theta$, $\cos\theta \approx 1$ and $\delta k \approx 2\sqrt{2}/r_0$. As detailed in the main text, for small angles, the coupling rate has a Gaussian dependence on phase mismatch, with a characteristic width given by the inverse of the optical beam size.

The dependence of the coupling rate on the angle of incidence can be derived by expanding the phase mismatch as

$$\Delta q = q_m - 2k_0 \sin\theta$$

$$\Delta q = 2k_0(\sin\theta_m - \sin\theta) \qquad (82)$$

Where the phase matching angle $\theta_m$ is defined as

$$\theta_m = \sin^{-1}\left(\frac{q_m}{2k_0}\right) \qquad (83)$$

For small angles of incidence θ, sin θ ≈ θ, the phase mismatch can be approximated as

$$\Delta q \approx 2k_0(\theta_m - \theta)$$

The dependence of the coupling rate can now be expressed as-

$$g_0 \propto \exp\left(-\frac{(\theta-\theta_m)^2}{\delta\theta^2}\right) \qquad (84)$$

where $\delta\theta = \frac{2\sqrt{a}}{2k_0} = \frac{\sqrt{2}}{k_0 r_0}$

For optical fields with free space wavelength of $\lambda_0 = 1550$ nm, and the optical beam size of $r_0 = 30$ μm, the angular bandwidth is $\delta\theta = 0.66°$, corresponding to a full width of $2\delta\theta = 1.32°$. Since the Brillouin gain coefficient, $G$ varies as the square of coupling rate- $G \sim g_0^2$, the corresponding angular bandwidth is given by $\delta\theta_G = \frac{1}{k_0 r_0} = 0.46°$ with a full width of $2\delta\theta_G = 0.92°$, agreeing well with the experimental results detailed in the main text.

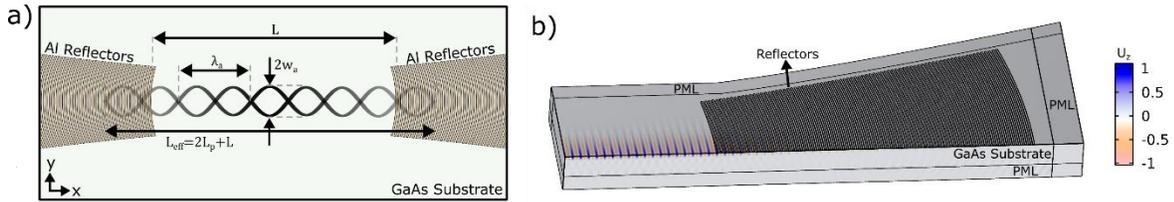

Figure S2: 3D FEM simulation of designed SAW cavities. a) Cross-sectional illustration of Gaussian SAW cavities. The two acoustic mirrors consisting of numerous Al reflectors, confine a standing wave SAW mode along the x-axis with a Gaussian profile along the y-axis. The illustration also displays independent design parameters that characterize the cavity, namely, the acoustic wavelength ($\lambda_a$), cavity orientation relative to the crystallographic axes (x-axis), acoustic waist ($w_a$) and mirror separation ($L$). b) Example of a COMSOL simulation of the the fundamental Gaussian mode of a SAW cavity along [100]-directions on [100]-cut GaAs. Parameters of the simulation are $\lambda_a = 5.7\ \mu m$, $f_a \sim 465\ MHz$, $L \sim 100\ \mu m$, $w_a = 3\lambda_a$ and reflector thickness $t = 0.035\lambda_a$.

## S3. Design and simulation of Gaussian SAW cavities

Gaussian SAW cavities in this work are based on a Fabry-Perot cavity design where two acoustic Bragg mirrors, each consisting of numerous metallic strips, confine surface acoustic modes in the region enclosed between them. Each metallic strip reflects a small portion of the incident acoustic field, and the cumulative interference by all the reflectors achieves the desired acoustic confinement. The geometry of the cavity is specified by four independent parameters- the direction of the cavity axis relative to a principal crystal axis (shown along the x-axis in Fig. S2a), acoustic wavelength ($\lambda_a$), the acoustic beam waist at the center of the cavity ($w_a$) and mirror separation ($L$) (Fig. S2a). For the chosen cavity axis, the acoustic group velocity ($v_g(\Theta)$) is calculated as a function of the angle relative to the axis ($\Theta$) by numerically solving acoustic wave equations with appropriate boundary conditions[10]. Accounting for this anisotropy of the SAW velocity of the underlying substrate is essential to designing an efficient SAW cavity. For a Gaussian beam with a beam waist $w_a$ and wavelength $\lambda_a$ the acoustic Rayleigh range ($x_R$) is given by $x_R = \pi w_a^2/2\lambda_a$ and the corresponding phase along the propagation axis (x-axis) can be expressed as $\Phi(x) = k_a x + \frac{1}{2}\tan^{-1}(x/x_R)$. The first and second terms refer to the propagation and Guoy phases. The locations of the reflectors ($x_i$) can now be determined by calculating the nodes of the acoustic displacement, i.e., $\Phi(x_i) = n\pi$. The location of the reflector closest to the center ($x_1$) is chosen such that the separation between the first reflector of the two mirrors is approximately $L$, i.e. $x_1 \approx L/2$. The mirror separation $L$ is chosen to be large enough to accommodate the optical beams incident on the device with minimal optical overlap with the acoustic mirrors. For efficient confinement of the acoustic field, the curvature of each metallic reflector must coincide with the local phase front of the desired Gaussian SAW mode. Ignoring the anisotropy in SAW velocity, the local phase front of a Gaussian beam at any reflector location $x_i$ would be circular arcs with a radius of curvature given by $R(x_i) = x_i\left(1 + \left(\frac{x_R}{x_i}\right)^2\right)$. To account for the anisotropy of the substrate, a correction factor[11] $v_g(\Theta)/v_g(0)$ is introduced to obtain an angle dependent radius of curvature function $R'(x_i, \Theta) = R(x_i)v_g(\Theta)/v_g(0)$.

To validate the design principles, we perform 3D numerical finite element simulations (COMSOL 5.6) of SAW cavities on [100]-cut GaAs oriented along [100]-direction with $\lambda_a = 5.7\ \mu m$, $L \sim 100\ \mu m$ and $w_0 = 3\lambda_a$. The thickenss of the metallic reflectors is specified as a fraction of the acosutic wavelength and is set to be $\frac{t}{\lambda_a} = 0.035$. This thickness found to be a good balance between achieving tight confinement and achieve small acoustic mode volumes (thick electrodes) and mitigate acoustic scattering into the bulk of the substrate (thin electrodes). This value of reflector thickness is in agreement with other similar works[12,13]. To minimize computational resources required, we leverage the symmetry in device and simulate one-fourth of the entire device. The FEM geometry of the device (Fig. S2b) consists of a substrate with a thickness of $3\lambda_a$ is surrounded by phase matched layers with thickness of $2\lambda_a$. In all areas of the device, the mesh size is ensured to be less than or equal to $\lambda_a/4$. Simulated devices only have 50 as opposed to 200 in the fabricated devices to limit computational resources required for the simulation. Given the large number of nodes in the simulation, the simulations are run on a super computing cluster with 56 nodes and 500-1000 GB RAM.

## S4. Device Fabrication

SAW resonator designs are translated on GaAs substrate via a standard e-beam lithography process (Fig. S3). First, we coat a double-side-polished GaAs chip with ~ 500 nm thick PMMA polymer.

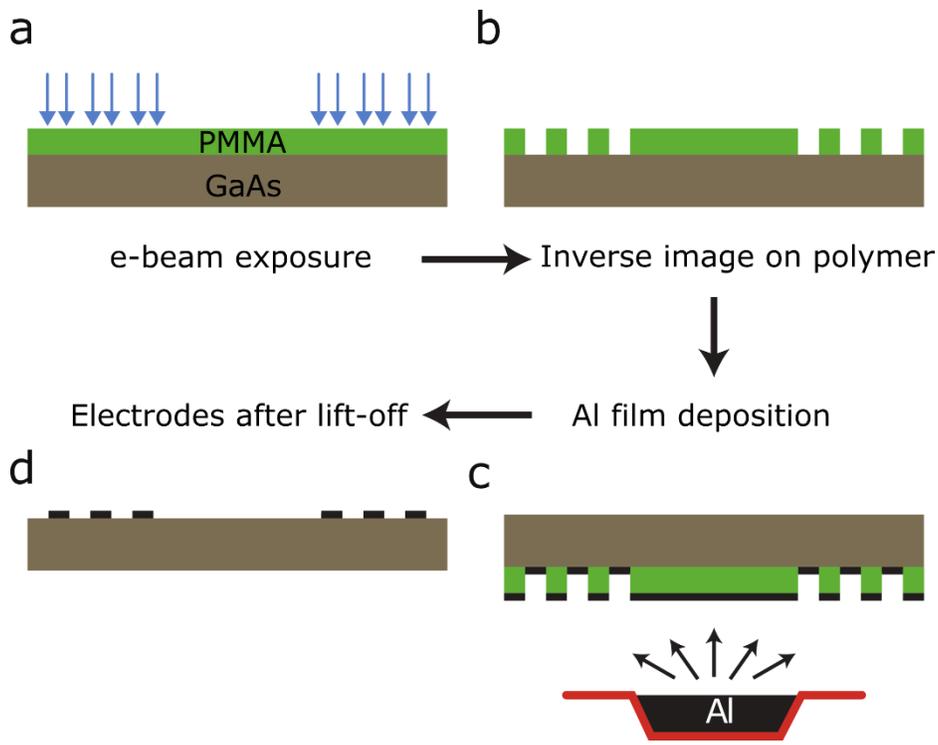

Figure S3: a. Drawing pattern on a PMMA-coated GaAs chip using a beam of electrons. b. Inverse image of pattern after development of the polymer in a 3:1 mixture of MIBK and IPA. c. Al film deposition in a UHV e-beam evaporation system. d. Al electrodes on GaAs after the lift-off.

The design is drawn onto the polymer with a beam of electrons using an electron-beam lithography tool (Fig. S3a). In the subsequent step, the polymer broken by e-beam exposure is washed away, resulting in a negative image of the pattern (Fig. S3b). Next, 200 nm thick Al film is deposited on the chip in an ultra-high vacuum e-beam evaporation system (Fig. S3c). Finally, the chip is removed from the chamber and submerged in a hot acetone bath, which removes PMMA and metal film from unwanted areas, leaving behind Al electrodes (Fig. S3d).

## S5. Optomechanical Spectroscopy Setup

This section presents additional details for the experimental spectroscopy apparatus used for measuring the Optomechanical response from SAWs (Fig. S4a). A continuous-wave (CW) laser at 1550 nm (the carrier, $\omega_C$) is divided into two fiber paths. Along one path, the optical field is modulated by a null-biased intensity modulator with a fixed frequency of $\omega_1 = 2\pi \times 11$ GHz, followed by a narrow fiber Bragg grating (FBG) to filter out the upshifted optical frequency sideband. The remaining lower frequency optical sideband serves as one of the acoustic drive tones, drive₁, with frequency $\omega_{d1} = \omega_C - \omega_1$. Similarly, along the second path, the carrier is modulated with a frequency of $\omega_2 = 2\pi \times (11 + \Omega)$ GHz and subsequently filtered with an FBG to generate the second acoustic drive, drive₂, with frequency $\omega_{d2} = \omega_c + \omega_2$. $\omega_{d1}$ and $\omega_{d2}$ are chosen such that the difference between the two ($\Omega = \omega_{d2} - \omega_{d1}$) can be continuously varied through targeted acoustic resonance frequencies (Fig.S4b).

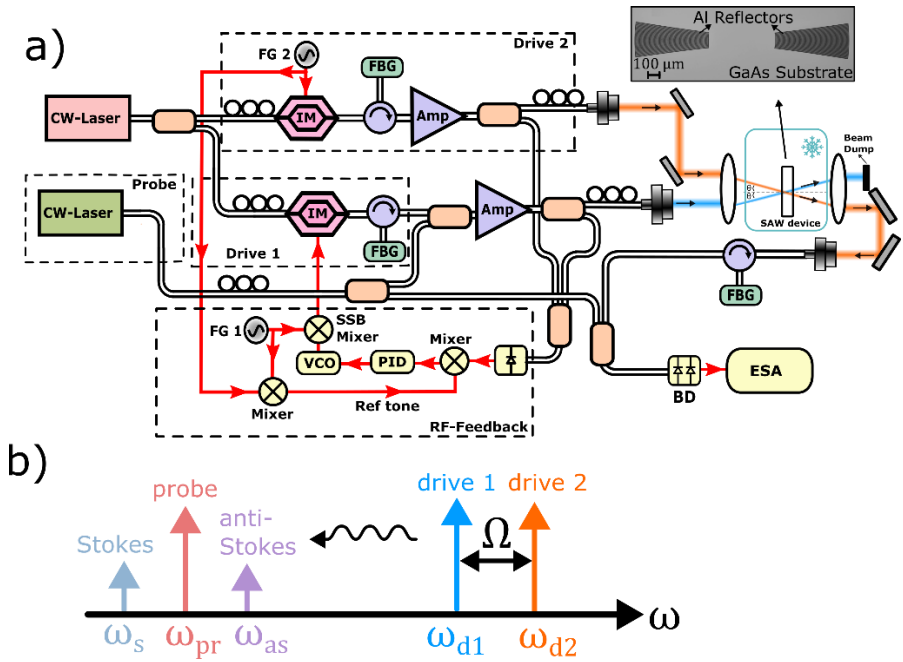

Figure S4: a) Experimental apparatus. IM: intensity modulator, FBG: fiber-bragg grating, Amp: optical amplifier, BD: balanced detector, ESA: electrical spectrum analyzer. b) Frequencies of the relevant optical tones. The acoustic wave with frequency $\Omega$ is optically excited by two acoustic drives (drive₁ at $\omega_{d1}$ and drive₂ at $\omega_{d2}$), and the Stokes ($\omega_S$) and anti-Stokes ($\omega_{AS}$) sidebands are generated on either side of the probe ($\omega_{pr}$) with offset, $\Omega$.

A second CW laser with a frequency $\omega_{pr}$ which serves as both the probe and the local oscillator (LO), is split into two optical paths. The optical field along the first path, which serves as the optical probe field that scatters off the driven acoustic field, is combined with drive₁. The optical field along the second path is combined with the final signal and serves as a local oscillator (LO) to detect optomechanically scattered signals. The three optical fields (drive₁, drive₂, and probe) are amplified by Erbium optical amplifiers to optical powers of about $\sim 150 - 350$ mW each before impinging on the SAW cavity. Two sets of polarization controllers ensure that the optical fields are linearly polarized. The optical fields are collimated and incident off-axis focusing aspheric lens. The off-axis displacement is controlled through a linear stage, providing fine control over the incident angle. The dependence of the incident angle on the off-axis displacement is carefully pre-calibrated, as detailed in the following section. The incident probe, which is collinear to drive₁, optomechanically scatters via the driven SAW cavity mode into a signal collinear to drive₂. This optomechanical signal and drive₂ are collected with a single mode collimator and filtered with an

FBG filter to reject the excess acoustic drive. The filtered signal is combined with the local oscillator and detected on a balanced detector.

The two acoustic drive tones (drive$_1$, drive$_2$) travel through distinct optical paths with different lengths and components. As a result, the optical path length between the two drive fields could vary by as much as tens of meters of single-mode fiber. The uncorrelated noise along the two optical paths as a result of environmental fluctuations (vibrations, temperature, etc.) could lead to excess relative optical frequency noise and limit the measurable linewidths of the optomechanical response. To mitigate the effects of the noise, we implement an optoelectronic feedback loop to lock the two acoustic drive tones. A small fraction (~ 1%) of the two acoustic drives are extracted before they are incident on the device under test and mixed on a photodetector. The phase of the resultant signal is measured on a lock-in amplifier, and a PID produces a correction signal to a voltage-controlled oscillator (VCO) for feedback. This phase-locked loop ensures that the two drive tones are stabilized to a relative frequency of ~ 1 Hz.

## S6. Calibrating Optical Angle of Incidence

In the paraxial limit, the off-axis optical rays axis passing through an ideal lens with a focal length of f intersects the optical axis at the focal point with an angle of incidence given by $\theta = \frac{s}{f}$, where s is the off-axis displacement (Fig. S5a). This elementary relation forms the basis of the angle-tuning technique employed in this work. This relation, in general, is not valid for real lenses with additional aberrations incident with diffracting optical beams. These effects (diffraction and aberrations) could result in significant deviations from the paraxial relation. To accurately determine the angle of incidence as a function of off-axis displacement, we develop an apparatus to image the focus of two intersecting beams and calculate the angle of incidence by observing the resulting interference pattern (Fig. S5b). An optical beam, named $beam_1$, is first incident along the optical axis of the lens under test, which subsequently focuses on a partially reflective sample placed at the focal plane of the lens. The focused beam is aligned to the surface normal of the sample by maximizing the back-reflected beam. This configuration is assumed to represent $\theta = 0°$. Next, $beam_1$ is laterally displaced by a known distance away from the optical axis. A second beam, $beam_2$, is aligned such that the partially reflected $beam_1$ maximally couples into the $beam_2$ collimator. This alignment ensures the two beams are focused on the same spot on the sample with equal but opposite angles of incidence. A 90:10 beam splitter samples a small portion of back reflected beams which are focused using an imaging lens on a near-infrared camera. The image observed on the camera consists of spatial fringes resulting from interference of $beam_1$ and $beam_2$ (inset Fig.S5c). The spatial periodicity ($\Lambda$) of the observed fringes can then be used to infer the angle of incidence on the sample through the relation $\Lambda = \frac{\lambda_0}{2 \sin\theta}$. The angle of incidence measured as a function of the off-axis displacement of $beam_1$ shows excellent agreement with predictions from paraxial theory (Fig. S4c) for an aspheric lens with a focal length of f = 75 mm. Observed results confirm that geometric aberrations within the lens and other optical components within the system are small, and paraxial analysis is warranted.

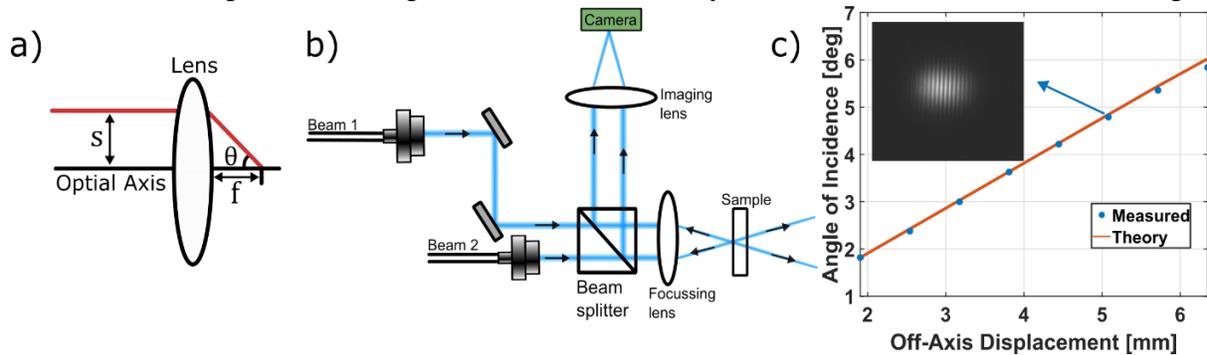

Figure S5: Measurement and caliberation of angle of incidence. a) An off-axis ray incident on a aberration free lens intersects the optical axis at the focal point with a angle given by $\theta = \frac{f}{s}$, b) Experimental apparatus used to measure the angle of incidence by imaging the focal plane, c) Measured angle as a function of off-axis displacement is compared to the ideal lens and displays good agreement. A typical image of the focal plane is inset.

## S7. Estimating Optomechanical Coupling Rate

This section discusses the theory of estimating the optomechanical coupling rate ($g_0$) from experimentally measured spectra. The system under consideration is as follows (Fig. S4b)- two optical drive fields with amplitudes $a_{d1}$ and $a_{d2}$ resonantly drive a SAW cavity mode with amplitude b. A third optical probe with an amplitude $a_{pr}$ scatters of the driven phonon mode and scatters into optomechanically scattered Stokes ($a_S$) and anti-Stokes sidebands ($a_{AS}$). Here two simplifying assumptions are made- first, the system is operated in the small gain limit, in which pump depletion does not occur and, as a result, incident optical fields $a_{d1}, a_{d2}$ and $a_{pr}$ do not evolve in space. Second, the weak phonon drive generated from the scattered signals ($a_S$ and $a_{AS}$) and the incident probe is neglected. In the strong gain limit, for instance, within a cavity optomechanical cavity, both of these assumptions would break down and require a more general analysis. Additionally, we assume that for small angles of incidence (near normal incidence), which is the case in this work, the optical beams approximately travel along the z-axis. In the limit of large angle this analysis can be suitably modified. The equations of motions for the driven cavity phonon amplitude (b) and scattered signals are given by $(a_S, a_{AS})^{3,14,15}$ –

$$\frac{\partial b}{\partial t} = -i(\Omega_0 - \Omega)b - \frac{\Gamma_0}{2}b - i \int g_0^* a_{d1}^* a_{d2} \tag{85}$$

$$v_g \frac{\partial a_S}{\partial z} + \frac{\partial a_S}{\partial t} = -ig_0^* b^* a_{pr} \tag{86}$$

$$v_g \frac{\partial a_{AS}}{\partial z} + \frac{\partial a_{AS}}{\partial t} = -ig_0 b a_{pr}^* \tag{87}$$

where $v_o, \Omega_0, \Omega, \Gamma$ refer to the optical group velocity, resonant phonon frequency, the frequency difference between optical drives, and acoustic dissipation rate. Note that the integral in Eqn. 85 is performed only over the interaction volume, i.e., the decay length of the SAW cavity mode within the bulk of the substrate, i.e. $l_a = \frac{1}{2\mathrm{Re}(\eta)q} \sim 2$ μm as defined in section S1. Assuming steady-state operation ($\partial_t = 0$) and resonant driving from optical drives ($\Omega = \Omega_0$) the phonon-field amplitudes can be expressed as-

$$b = -i\frac{2}{\Gamma}g_0^* l_a a_{d1}^* a_{d2} \tag{88}$$

Inserting Eqn. 47 into Eqn. 44 and Eqn. 45 and assuming $a_{S(AS)}(0) = 0$ one obtains-

$$a_S = \frac{2|g_0|^2 l_a^2}{\Gamma v_g} a_{d1} a_{d2}^* a_{pr} \tag{89}$$

$$a_{AS} = \frac{2|g_0|^2 l_a^2}{\Gamma v_g} a_{d1}^* a_{d2} a_{pr}^* \tag{90}$$

The optical power in terms of field amplitude is given by-

$$P_i^{op} = \hbar\omega_i v_g |a_i|^2 \tag{91}$$

Using Eqn. 48, 49 and 50, the optomechanically scattered sideband powers can be expressed as-

$$P_{AS} = P_S = \frac{\beta^2}{\hbar^2 \omega_{d1} \omega_{d2} v_g^2} P_{d1} P_{d2} P_{pr} \qquad (92)$$

Where $\beta = \frac{2|g_0|^2 l_a^2}{\Gamma v_g}$. These optomechanically scattered sidebands are spectrally separated by $\Omega_0$ on either side of the incident probe. Assuming a local oscillator with an optical power $P_{LO}$, the resulting heterodyne beat note oscillating at a frequency $\Omega_0$ is given by-

$$P_{het} = 2\sqrt{P_{LO} P_S} + 2\sqrt{P_{LO} P_{AS}} = 4\sqrt{P_{LO} P_{AS}} = \frac{8|g_0|^2 l_a^2}{\hbar \omega_{d1} \Gamma v_g^2} \sqrt{P_{d1} P_{d2} P_{pr} P_{LO}} \qquad (93)$$

The coupling rate can then be estimated by inverting equation Eqn. 93 as

$$g_0 = \left( \frac{P_{het} \hbar \omega_{d1} \Gamma v_g^2}{8 l_a^2 \sqrt{P_{d1} P_{d2} P_{pr} P_{LO}}} \right)^{\frac{1}{2}} \qquad (94)$$

**[100]-oriented cavities:**

For cavities oriented along the [100]-direction, optical drive tones, and the probe tone have a free space wavelength of $\lambda_p = 1550.05$ nm and $\lambda_{pr} = 1550.25$ nm, respectively. The optical fields are incident at an angle of 7.8°. Drive$_1$, Drive$_2$, and probe powers before the sample are 325 mW, 117 mW, and 434 mW, respectively. Optical reflectivity at the front surface resulting from refractive index mismatch is 29.41 %. Since reflected fields do not contribute to the optomechanical process, the effective powers mediating the optomechanical process are $P_{d1} = 228$ mW, $P_{d2} = 82$ mW and $P_{pr} = 304$ mW. All the optical fields are ensured to be TE-polarized (p-polarization) by using a polarizing beam splitter. The local oscillator power is $P_{LO} = 3.7$ mW. The measured heterodyne power is $P_{het} = 0.34$ µW. Corresponding to an experimentally estimated coupling rate of $\frac{g_0}{2\pi} = 1.43 \times 10^3$ Hz. This agrees well with a theoretically estimated value of $1.73 \times 10^3$ Hz. Residual errors could be a result of uncalibrated rf losses, polarization mismatch between the optical tones, and errors in optical beam position and sizes.

**[110]-oriented cavities**

For cavities oriented along the [110]-direction, optical drive tones, and the probe tone have a free space wavelength of $\lambda_p = 1550.05$ nm and $\lambda_{pr} = 1550.25$ nm, respectively. The optical fields are incident at an approximate angle of 7.8°. Drive$_1$, Drive$_2$, and probe powers before the sample are 332 mW, 126 mW, and 443 mW, respectively. Optical reflectivity at the front surface resulting from refractive index mismatch is 29.41 %. Since reflected fields do not contribute to the optomechanical process, the effective powers mediating the optomechanical process are $P_{d1} = 232$ mW, $P_{d2} = 88.7$ mW and $P_{pr} = 310$ mW. All the optical fields are ensured to be TE-polarized (p-polarization) by using a polarizing beam splitter. The local oscillator power is $P_{LO} = 3.7$ mW. The measured heterodyne power is $P_{het} = 21.8$ nW. Corresponding to an experimentally estimated coupling rate of $\frac{g_0}{2\pi} = 1.90 \times 10^3$ Hz. This agrees well with the theoretically estimated value of $1.8 \times 10^3$ Hz.

## S8. Quality factor vs. length

The quality factor of an acoustic cavity (Q) can be expressed as a function of roundtrip loss ($\alpha_l$), resonant frequency ($f_0$), acoustic velocity ($v_R$), cavity length ($L$), and linewidth ($\Delta f$) as[12]-

$$Q = \frac{f_0}{\Delta f} = \frac{2f_0 L}{v_R \alpha_l} \qquad (97)$$

The acoustic round trip loss can be expressed as a sum of the propagating loss and losses occurring in the acoustic mirror. Propagation losses which scale with propagation length, can be characterized through an attenuation coefficient ($\alpha_P$) while mirror losses ($\alpha_M$) are independent of length. The total loss can now be expressed as-

$$\alpha_l = 2(\alpha_p L + \alpha_M) \qquad (98)$$

The factor of two in Eqn. 98 is a result of acoustic fields propagating for a total round trip length of $2L$ and encounter the acoustic mirrors twice, one on each side of the cavity. Inserting Eqn. 98 into Eqn. 97 we get

$$Q = \frac{f_0 L}{v_R} \left( \frac{1}{\alpha_p L + \alpha_M} \right) \qquad (95)$$

For small cavity lengths, assuming $\alpha_P L \ll \alpha_M$, that is, losses are dominated by mirror losses, we get-

$$Q \approx \frac{f_0 L}{v_R \alpha_M} \qquad (100)$$

quality factor as observed in the main text, depends linearly on the cavity length.

## S9. Absorption-mediated optomechanical effects

In addition to optomechanical interactions enabled by nonlinear optical forces (photoelastic and radiation pressure), devices investigated in this work, by virtue of having metallic reflectors, also display optomechanical interactions mediated by absorption within the metallic reflectors. The mechanism of this interaction is as follows- the two drive fields separated by the resonant cavity frequency act as an intensity-modulated source which, when absorbed by the acoustic reflectors, excites SAWs due to thermo-elastic expansion. Subsequently, the optical probe field can scatter off the excited SAW cavity mode to produce an analogous optomechanical response[16,17]. Note that such processes do not require phase-matching of the optical drives which generate the acoustic fields since the acoustic fields are driven solely by time-modulated absorptive effects.

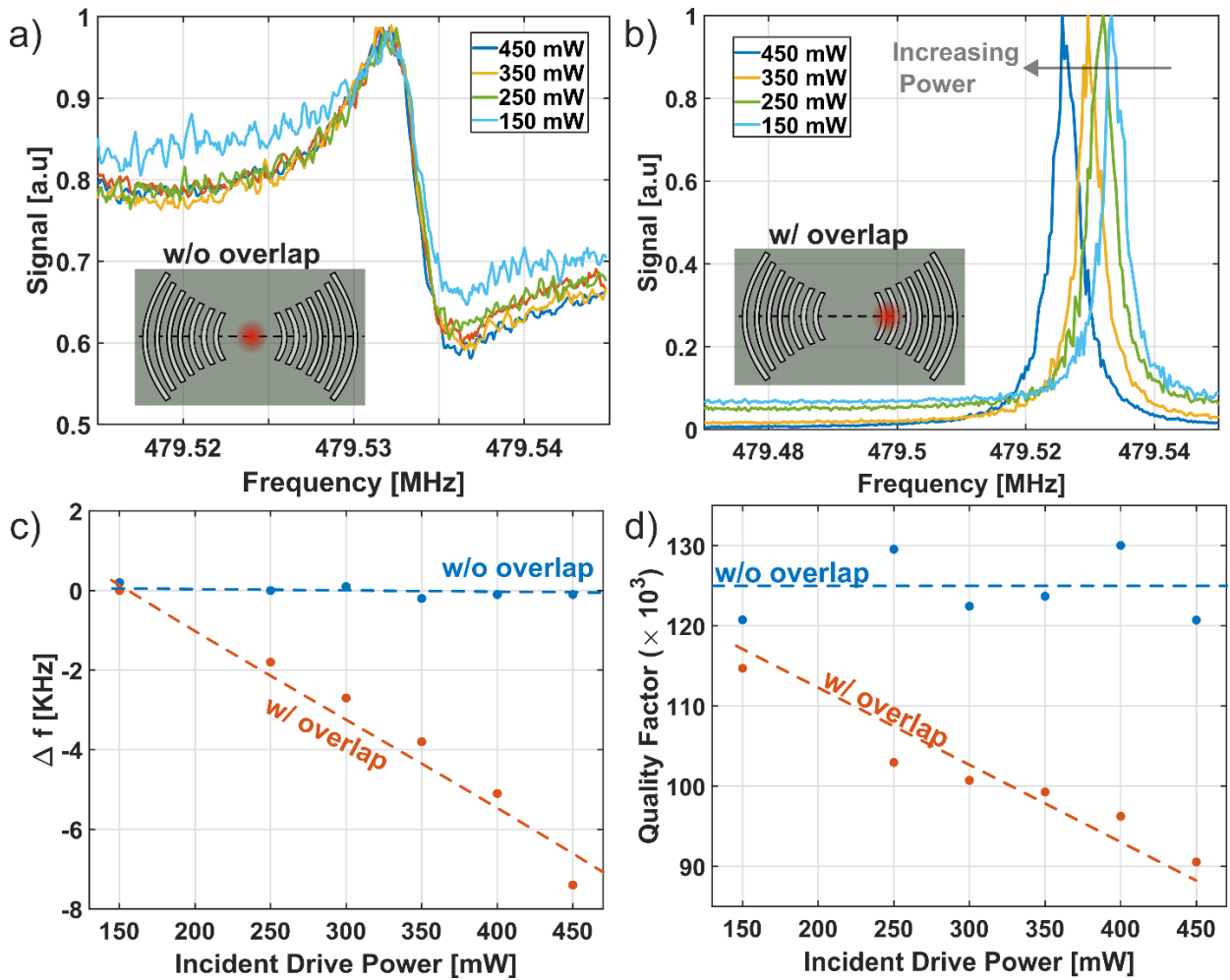

Figure S6. Absorption-mediated optomechanical processes. a) Optomechanical response as a function of incident drive power when the optical fields have minimal overlap with the metallic reflectors with illustration inset. b) Optomechanical response as a function of incident optical power when the optical fields overlap with the metallic reflectors (inset illustration). c) Change in the resonant frequency ($\Delta f$) of the observed cavity mode as a function of incident optical drive power. We define $\Delta f = 0$ at an incident drive power of 150 mW. d) Acoustic quality factor as a function of incident optical power with and without optical overlap with the metallic reflectors.

Since absorptive effects are typically accompanied by thermal effects, such residual thermal effects are used to discriminate the two optomechanical effects (parametric and absorptive). Optomechanical response in the [100]-oriented devices is measured as a function of incident optical power when the optical beams are in the center of the cavity (parametric) (Fig. S6a), and when the optical fields have significant overlap with the surrounding metallic reflectors (absorption-mediated) (Fig.S6b). For the case where parametric interactions dominate the optomechanical response (Fig. S6a), negligible power-dependent effects are observed (Fig. S6c-S6d). In stark contrast, the resonant frequency and the quality factor vary significantly as a function of incident power for the absorption-mediated response (Fig. S6b-d). The observed differences suggest that excess optical absorption in the metal strips modifies the characteristics of the resonant mode, consistent with spurious heating of the substrate and associated changes in local elastic properties of the SAW cavity. As demonstrated in previous works, such absorptive effects could be employed for various classical applications, including optical signal processing[16,17]. However, given their incoherent nature, absorptively mediated effects would generally be undesirable for applications requiring coherent interactions, including quantum control, transduction, and sensing. Additionally, spurious heating resulting from absorption could prevent the robust ground-state operation of quantum systems such as qubits. These parasitic thermal effects are minimized for devices investigated in this work by ensuring g the mirror separation is much larger than the incident optical beam waist. For example, for devices employed in this work with an approximate mirror separation of $500 \ \mu m$ and an optical beam of waist diameter of $60 \ \mu m$, the fraction of optical power spatially overlapping with the acoustic mirrors can be reduced to the level of $\sim 10^{-15}$. Alternatively, any residual thermal effects can be eliminated by replacing the metallic stripe reflectors with etched grooves to confine SAWs[13,18,19].

### S10. SAW mediated cavity optomechanical devices

Here, we propose an possible iteration of a SAW-mediated cavity optomechanical system. For the SAW cavity system we assume a SAW cavity on [100]-cut GaAs optimized for an acoustic wavelength (frequency) of $\lambda_a = 700$ nm ($\Omega_0 \sim 4$ GHz) with a Gaussian waist size of $w_0 = 2\lambda_a$ and a cavity length of $L_{\text{eff}} \sim 30\lambda_a$. This SAW cavity can be phase matched to $\lambda_o = 1$ $\mu m$ optical fields incident at $\theta = 45°$. Assuming the fields are TM polarized Eqn. 78 can be used to estimate traveling-wave coupling rate of $g_0 \sim 2\pi \times 400$ $kHz$. This estimated coupling rate is approximately 250x of experimentally measured cavities which had a frequency of 500 MHz. This large enhancement is a result of the acoustic mode volume scaling with the wavelength.

Next we propose a possible optical cavity compatible with SAW cavities. Consider a coated DBR fiber-optic optical cavity enclosing a 4-GHz SAW cavity. Optical cavities like these have been commonly used membrane-type cavity optomechanical systems and CQED systems[20,21]. The ability to miniaturize these optical cavities is ideal to obtain small optical mode volumes and consequently larger cavity optomechanical coupling strengths. Assuming optical cavity lengths of tens of micron ($l_{opt} \sim 10 - 100$ $\mu m$) and a conservative optical finesse of $\mathcal{F} = 10^4$ (these could be as large as $10^6$). Using Eqn. 8 we estimate the cavity optomechanical coupling rate as (for $l_{opt} = 10$ $\mu m$)-

$$g_0^c = g_0 \left(\frac{l_a}{l_{opt}}\right) \approx 2\pi \times 5.5 \; kHz \tag{101}$$

The reduction in coupling rate when compared to the travelling-wave coupling rate quoted previously ($g_0 \approx 2\pi \times 400$ $kHz$) results from the relative modal size mismatch of the acoustic ($\sim 0.5$ $\mu m$) and optical cavity ($\sim 10 - 100$ $\mu m$) modes. The extraordinary power handing capabilities of this system, limited only by bulk material thresholds could support large intracavity photon numbers ($n_c > 10^9$), typically employed in Bulk cavity optomechanical systems[22]. The optically loaded cavity optomechanical cooperativity is given by[4]

$$C_{\text{om}} = \frac{4g_0^2 n_c}{\Gamma_a \kappa} \tag{102}$$

Where $\Gamma_a$ and $\kappa$ refer to acosutic and optical cavity decay rates. Assuming experimentally observed quality factor $Q \approx 10^5$ and $g_0^c \approx 2\pi \times 5.5$ kHz the cavity optomechanical cooperativity is estimated to be

$$C_{\text{om}} \approx 2500 \tag{103}$$

This platform retains the high-power handling capability of bulk optomechanical systems[5,23] while offering much larger coupling rates (50-500x), devices with a smaller footprint, and simpler integrability to other quantum systems and sensing devices.

**S11. Additional experimental data**

Here additional experimental data is presented for the [100]-oriented device on [100]-cut GaAs (Fig. S7) for the cases where one of the acoustic drives is turned off (green trace Fig. S7a), when the two acoustic drives are orthogonally polarized with respect to each other (TE-TM scattering, purple trace Fig. S7a), and when the optical LO is polarized orthogonally to the probe field (yellow trace Fig. S7a). These results are in excellent agreement with theoretical predictions and strongly suggest that the observed resonance results from optomechanical processes. The TM-TM scattering trace is also shown (Fig. S7b), displaying a resonance with an estimated quality factor of 120,000.

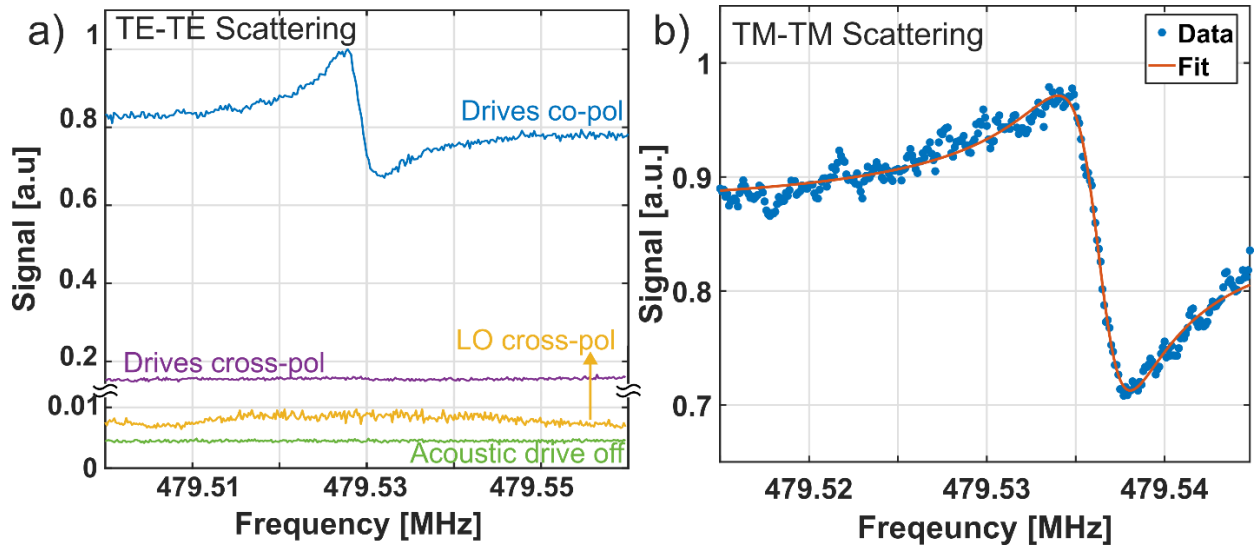

Figure S7: Additional optomechanical measurement data. a) Additional TE-TE scattering data including when the acoustic drives are turned off (green), the two acoustic drives are orthogonally polarized with respect to each other (purple) and when LO is orthogonally polarized (yellow) with respect to incident probe. TE-TE scattering response is also shown for reference (blue). b) Optomechanical response when the all the fields are TM polarized.